\documentclass[aps,pra, longbibliography,twocolumn,nofootinbib]{revtex4-2}
\usepackage{amsmath}
\usepackage{amssymb}
\usepackage{graphicx}
\usepackage{comment}
\usepackage{soul,xcolor}
\usepackage{bbold}
\usepackage[colorlinks=true ,urlcolor=blue,urlbordercolor={0 1 1}]{hyperref}
\usepackage{tikz}
\usepackage{braket}
\usepackage[english]{babel}
\usepackage{graphicx}
\usepackage{lipsum}
\usepackage{xr}
\usepackage{pifont}
\usepackage{multirow}
\usepackage{ulem}

\newcommand{\del}{\nabla}

\newcommand{\be}{\begin{equation}}
\newcommand{\ba}{\begin{align}}
\newcommand{\ee}{\end{equation}}
\newcommand{\bea}{\begin{eqnarray}}
\newcommand{\eea}{\end{eqnarray}}
\newcommand{\beq}{\begin{equation}}
\newcommand{\eeq}{\end{equation}}
\newcommand{\beqn}{\begin{eqnarray}}
\newcommand{\eeqn}{\end{eqnarray}}

\newcommand{\bfk}{\mathbf{k}}
\newcommand{\bfq}{\mathbf{q}}
\newcommand{\bfr}{\mathbf{r}}
\newcommand{\bfl}{\mathbf{l}}

\newcommand{\cala}{\mathcal{A}}

\newcommand{\moire}{moir\'e }

\begin{document}

\title{Family of multilayer graphene superconductors with tunable chirality: Momentum-space vortices nucleated by a ring of Berry curvature
}
 
\author{Adarsh S. Patri}
\author{Marcel Franz}

 \affiliation{Department of Physics and Astronomy \& Stewart Blusson Quantum Matter Institute, University of British Columbia, Vancouver BC, Canada, V6T 1Z4}

\date{\today}

\begin{abstract}
Recent experiments in rhombohedrally-stacked multilayer graphene heterostructures have reported signatures of chiral superconductivity, emerging from a spin and valley-polarized normal state with broken time-reversal symmetry and an associated anomalous Hall effect. These findings bring into focus the role of the electronic Bloch wavefunction and the quantum geometric tensor in determining the superconducting pairing channel.
In this work, we examine superconducting instabilities of a model of $N$-layer rhombohedral graphene that possesses an enhanced Berry curvature distribution on an extended ring in momentum space -- that we dub the `Berry ring of fire' -- in the presence of an isotropic attractive interaction with a parametrically controlled spatial range.
We determine that local interactions favor a $N$-fold winding in the order parameter phase for odd-$N$ layered systems, with even-$N$ layers requiring a spatially extended attraction range to achieve pairing. 
For generic interaction lengths, we discover a family of chiral superconductors and, remarkably, momentum-space vortices nucleated on the Berry ring of fire. The existence of these vortices can be traced to a momentum-space flux quantization condition involving the Berry curvature, with the phase winding dictated by a combination of the Berry flux and a `statistical flux' to enforce Fermi-Dirac statistics. 
Such an order parameter structure allows for the possibility of {\it in-situ} tuning between various chiral superconducting phases through changes in the electron density or the displacement field. We discuss ways in which these predictions can be experimentally tested and potentially exploited in future devices.
\end{abstract}
\maketitle

\section{Introduction}

The confluence of strong correlations and topology in tunable two-dimensional electronic systems presents the exciting opportunity 
to re-examine and re-imagine the phenomenology of macroscopic quantum phenomena. 
This undertaking has been partly driven by recent quantum anomalous Hall experiments in two-dimensional \moire heterostructures such as twisted transition metal dichalcogenides (TMDs)~\cite{cai2023signatures, zeng2023integer,park2023observation,xu2023observation} and rhombohedral-stacked multilayer graphene aligned with hexagonal Boron Nitride (hBN)~\cite{lu2024fractional,lu2025extended}.
In these platforms, the necessary ingredients as well as the phenomenology of the quantum Hall effect have been generalized from the archetypal paradigm: a liquid state of matter arising from a pristinely flat Landau level with a uniform Berry curvature due to strong magnetic fields \cite{jain2007composite}.
Instead, quantum Hall phenomenology can occur at zero magnetic field where the electronic bandwidth is non-negligible, quantum geometric quantities are non-uniform \cite{goldman2023zero,dong2023composite, dong2024theory,dong2024anomalous,zhou2024fractional,kwan2023moir,soejima2024anomalous,dong2024stability,patri2024extended}, and can possibly even realize an electronic topological crystal known as the Hall crystal \cite{tevsanovic1989hall,zeng2024sublattice,soejima2024anomalous,dong2024stability,patri2024extended,desrochers2025elastic,soejima2025jellium, dong2025phonons}.
Very recently, in rhombohedral tetra-, penta- \cite{han2025signatureschiralsuperconductivityrhombohedral}, and hexa- \cite{morissette2025superconductivityanomaloushalleffect} layer graphene unaligned with hBN, in the same region of the density-displacement field tuned phase diagram where the aforementioned quantum anomalous Hall states were realized, unconventional chiral superconductivity has been discovered.
This discovery provides a unique opportunity to examine superconductivity emerging from topological bands, and to scrutinize the role of the Bloch wavefunction and quantum geometry in dictating the condensate.

\begin{figure}[t]
\includegraphics[width = 0.42\textwidth]{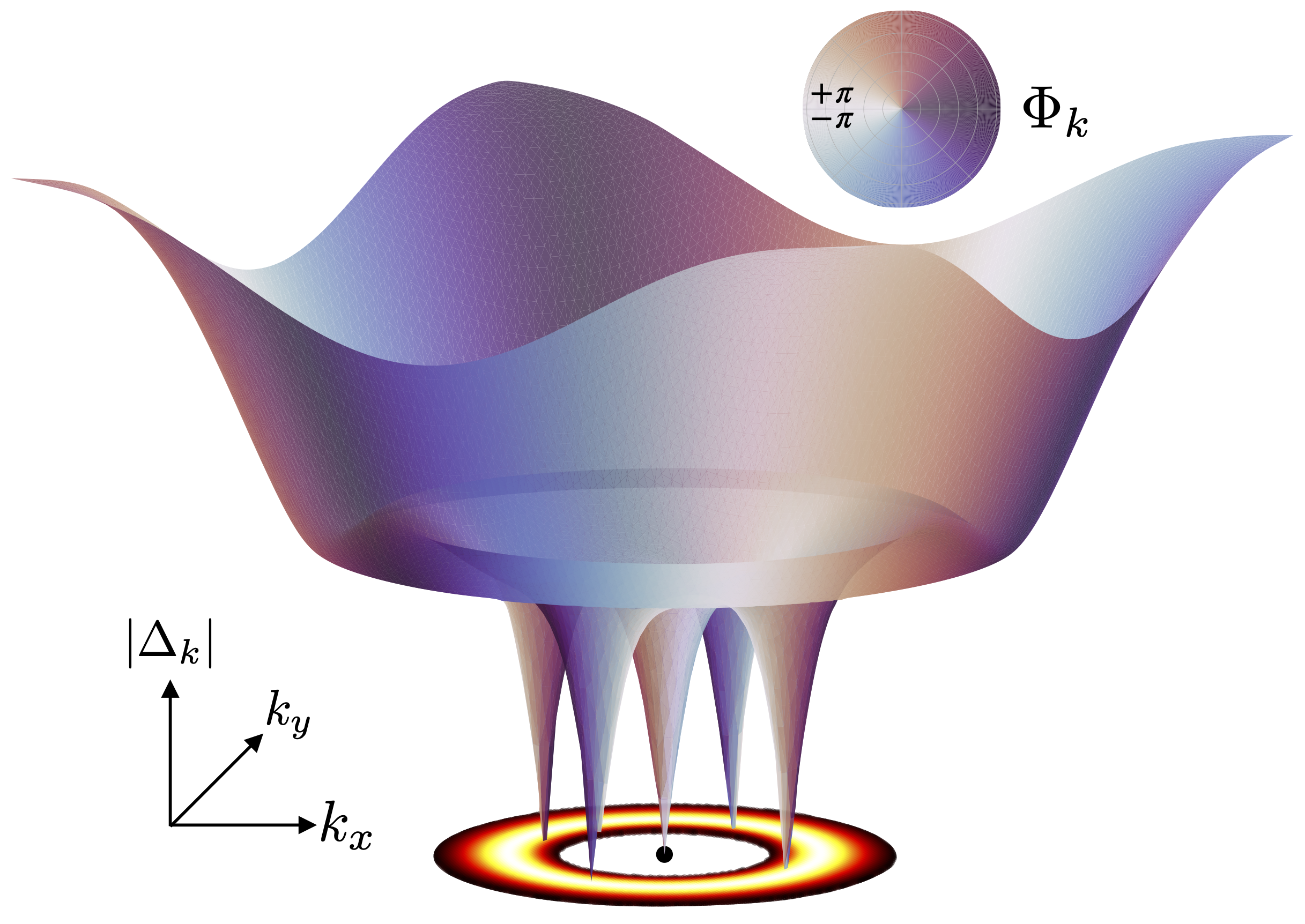}
\caption{Nucleation of momentum-space vortices on the Berry ring of fire (BRF). 
{To illustrate the central idea of this work, we depict} the magnitude of the superconducting order parameter $\Delta_k$ {being} suppressed at four points on the BRF (yellow disk), with a superconducting phase $\Phi_k$ winding by $\pm2\pi$ around these points, indicated by the color gradient.
The vortex at the origin (black dot) is the remnant from a momentum-isotropic bare attraction potential and reflects the odd spatial parity Cooper pair wave function, $p_x+ip_y$ in this case, required for pairs composed of same-spin electrons.}
\label{fig_main_bef_vortices}
\end{figure}
Superconductivity in non-\moire rhombohedral stacking of $N$-layers of graphene (R$N$G) had previously been observed in bilayer (R2G \cite{zhou2022isospin, zhang2023enhanced, holleis2025nematicity, li2024tunable, zhang2024twist}), trilayer (R3G \cite{zhou2021superconductivity,yang2408diverse, patterson2024superconductivity}), and tetralayer (R4G \cite{choi2024electric}) graphene.
These systems are generically susceptible to electronic instabilities stemming from their narrow-bandwidth kinetic energy.
The superconducting phenomenology in these platforms, though rich, ultimately involves pairing of electrons in opposite valleys ($K$ and $K'$) so as to satisfy time-reversal symmetry and render a Cooper pair of vanishing center-of-mass momentum.
On the other hand, in the afore-described samples of R4G and R5G~\cite{han2025signatureschiralsuperconductivityrhombohedral}, the normal state is a spin and valley polarized quarter-metal with an associated anomalous Hall effect.
This finite-momentum pairing superconductor emerging from a time-reversal broken normal state is reminiscent of the Fulde–Ferrell–Larkin–Ovchinnikov (FFLO) state \cite{casalbuoni2004inhomogeneous}, where here the Cooper pair carries a finite center-of-mass momentum at monolayer graphene valley $K$.

The recent breakthroughs in R4G and R5G have triggered a flurry of theoretical studies that have examined Kohn-Luttinger mechanisms of pairing (and the pairing symmetry of the order parameter)~\cite{chou2024intravalley, geier2024chiral, yang2024topological, jahin2024enhanced, qin2024chiral, yoon2025quarter, parramartinez2025bandrenormalizationquartermetals,christos2025finite,wang2024chiralsuperconductivityparentchern}; as well as more exotic origins of the superconductor, for instance by doping fractionalized states of matter~\cite{kim2025topological, shi2024doping,divic2024anyon}.
There have also been investigations of quantum geometry and its role in stabilizing a chiral superconductor within the context of toy models~\cite{shavit2024quantum, may2025pairing, dong2025controllabletheorysuperconductivitystrong}.
Currently lacking, however, is a universal examination of the family of superconductors that can emerge from electronic systems where there is significant Berry curvature and quantum metric in momentum space, as well as a careful disentanglement of the role of the quantum geometric tensor and the complete Bloch wavefunction of the normal state electrons. 

In this work, we provide an insight into the role of the Bloch wavefunction and quantum geometric quantities in determining the superconducting order parameter.
We examine the family of valley-polarized R$N$G within the context of a minimal two-orbital model.
For small momenta, a flat dispersion is generated by the applied perpendicular displacement field; while for larger momenta the $k^N$ dispersion of R$N$G dominates.
At the momentum scale associated with this band-dispersion crossover is a significant Berry flux enclosed within a ring in momentum space, as depicted schematically by the yellow ring in Fig.\ \ref{fig_main_bef_vortices}.
This region of intense Berry flux, which we dub a `Berry ring of fire' (BRF henceforth), provides the backdrop for us to examine the influence of the Berry curvature on the magnitude and phase of the superconducting order parameter.
We couch our study within a BCS-like treatment with an attractive interaction that we model by a Gaussian function with a real-space length $\alpha$.
This agnosticism to the pairing mechanism allows us to tune the spatial range of the interactions and investigate the different regimes when the entire Bloch wavefunction or quantum geometric tensor play a crucial role.

We discover that for odd-$N$ layer stacking of graphene, electron pairing is natural for contact attraction ($\alpha=0$, henceforth referred to as short-range interaction), and analytically demonstrate the chiral nature of the order parameter with $N$-fold winding of the superconducting phase.
For even-$N$ layers, local attraction does not give rise to pairing as it is forbidden by Fermi-Dirac statistics for electrons of same spin and valley; pairing occurs only for $\alpha>0$.
For such extended interactions we discover a family of chiral superconductors for both odd- and even-$N$ layers characterized by various phase windings, with the winding number undergoing an abrupt change at the BRF. This is facilitated by the nucleation at the BRF of momentum-space vortices in the superconducting order parameter depicted schematically in Fig.\ \ref{fig_main_bef_vortices}. Our results thus reveal the existence of a novel momentum-space vortex lattice whose existence is driven by the Berry curvature in analogy to the real-space Abrikosov vortex lattice driven by applied magnetic field.
In the analytically controllable limit of large $\alpha$, we indeed demonstrate that the origin of these momentum-space vortices can be traced to a momentum-space equivalent flux quantization condition wherein the Berry curvature takes the role of the magnetic field. The phase winding is determined by the sum of twice the enclosed Berry flux  (rounded to multiple of $\pi$) and a `statistical flux' (integer multiple of $2\pi$) to enforce Fermi-Dirac statistics and the odd-spatial parity of the gap function.
These different possible phase windings in momentum space suggest that tuning the size of the BRF (by displacement field) or size of the Fermi surface (by chemical potential) can lead to the realization of a family of distinct chiral superconductors with different Chern numbers.
These distinct chiral superconductors are separated by a quantum phase transition point, where the superconducting gap closes to yield a nodal superconductor, with associated experimental signatures.

The remainder of the manuscript is organized as follows.
In Sec.\ \ref{sec_microscopics} we introduce the microscopic model for rhombohedrally-stacked multilayer graphene and highlight the properties of the quantum geometric tensor and BRF.
We next discuss the BCS pairing attraction in the Bloch-projected basis and the symmetry-classification of the superconducting order parameter in Sec.\ \ref{sec_bcs_pairing_setup}.
In Sec.\ \ref{sec_local_sol} we present the chiral (broken time-reversal and odd-parity) pairing channels for local attraction in odd-$N$ layers. 
We next study spatially extended interactions (parameterized by $\alpha$) and find a family of chiral superconductors for both odd and even $N$-layers in Sec.\  \ref{sec_extended_int} with momentum-space vortices in the order parameter. 
This is followed by an analytical treatment of the small and large $\alpha$ limits in subsections \ref{sec_sub_alpha0} and \ref{sec_sub_alphainf}, where the origin of the momentum-space vortices and role of the Berry and statistical flux are also addressed.
In Section \ref{sec_transition_chiral_sc} we discuss density and displacement field tuned transitions between different chiral superconductors, and associated experimental signatures in Sec.\ \ref{sec_expt_signatures}. Finally, we conclude and propose directions of future work in Sec.\ \ref{sec_discussion}.

\section{Microscopic model of multilayer graphene}
\label{sec_microscopics}

\begin{figure}[t]
\includegraphics[width = 0.45\textwidth]{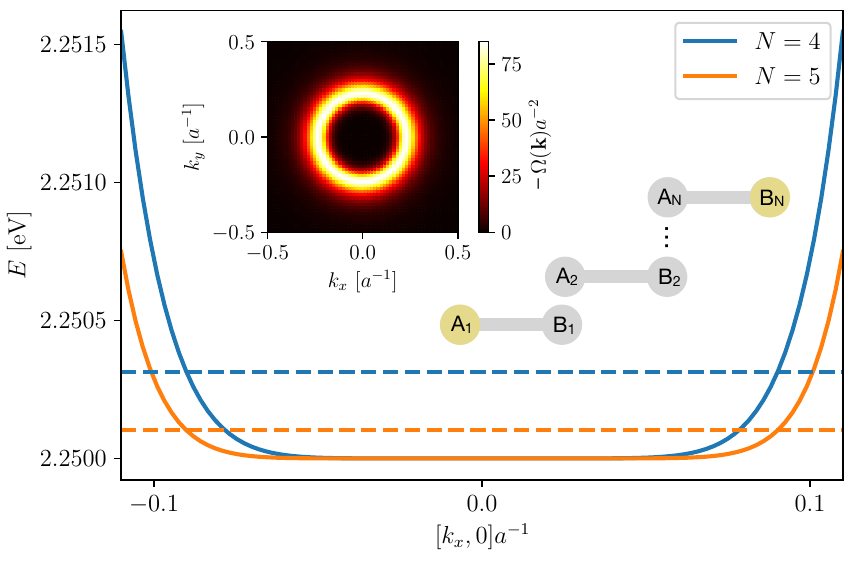}
\caption{Continuum conduction band dispersion for spin-valley ($\uparrow K$) polarized electrons in rhombohedral stacking (see right inset) of $N=4,5$ (blue and orange) layers of graphene. 
Fermi level (dashed line) is set such that a density of $10^{12}$ cm$^{-2}$ is filled about charge neutrality. Displacement energy of $|D| = 2.25$eV for illustration. 
Inset (Left): Berry curvature distribution in momentum space for $N=5$. The ring deforms into a $C_3$-symmetric shape with more realistic microscopic modelling \cite{dong2024stability}.}
\label{fig_band_berry_all}
\end{figure}

The two-orbital\footnote{The two-orbital model misses band structure properties about $\bfk=0$ that the full $2N$ orbital model includes: a $\pi$-Berry phase and negative effective mass for the kinetic energy~\cite{dong2024stability} that is required for more careful microscopic studies (such as that of quantum anomalous Hall effects, see Refs. \cite{dong2024theory,dong2024anomalous,zhou2024fractional,dong2024stability}).} continuum model of rhombohedral-stacked N-layer graphene (R$N$G) is a useful framework to examine the interplay of correlation effects and quantum geometry.
We consider the low-energy Hilbert space of electrons residing on the $A_1$ and $B_N$ sites, as depicted in the inset of Fig.\ \ref{fig_band_berry_all}.
The kinetic Hamiltonian in this $\{A_1, B_N\}$ orbital space for valley $K$ is of the form \cite{macdonald_trilayer_2010,zhang2019nearly}
\begin{align}\label{e1}
    H_0 = 
    \begin{pmatrix}    
    {-}D & v_N(k_x - i k_y)^N \\
    v_N(k_x + i k_y)^N & {+}D 
    \end{pmatrix}.
\end{align}
Here $D$ denotes the displacement field energy applied in the direction perpendicular to the multilayer stack, and $\bfk$ is the low-lying momentum of the generalized-Dirac cone (in units of inverse lattice constant $a^{-1}$), and $v_N$ is the generalized Dirac velocity (see Appendix \ref{app_miscroscopic_params} for microscopic parameters).
For the opposite valley $K'$, {the kinetic Hamiltonian is $[H_0(-\bfk)]^*$} from time-reversal symmetry.
The continuum band dispersion (depicted in Fig.\ \ref{fig_band_berry_all} for $N=4,5$) is of the form
\begin{align}
\epsilon_{\pm} (\bfk) = \pm \sqrt{v_N^{2}\bfk^{2N} + D^2}.    
\label{eq_kinetic_energy}
\end{align}

Motivated by the recent observation of chiral superconductivity in tetralayer ($N=4$) and pentalayer ($N=5$) graphene \cite{han2025signatureschiralsuperconductivityrhombohedral}, where the normal state is a quarter-metal of electrons (i.e., one that is spin-valley polarized), we primarily focus on the conduction ($+$) band electrons in valley $K$ and spin-$\uparrow$.
As seen in Fig.\ \ref{fig_band_berry_all}, the conduction band is increasingly flatter with increasing $N$ and displacement field energy $D$.
The subsequent enhanced density of states primes the normal state to a pairing instability in the presence of an attractive interaction.

Associated with this flat conduction band dispersion is the electronic Bloch wavefunction,
\begin{align}
    \ket{u (\bfk)} = \begin{pmatrix} a_\bfk \\ b_\bfk
    \end{pmatrix} = 
    \mathcal{C}_\bfk
    \begin{pmatrix}
     v_N(k_x - i k_y)^N \\ D + \sqrt{v_{N}^2\bfk^{2N} + D^2}
\end{pmatrix} ,
\label{eq_eigenbasis}
\end{align}
where $\mathcal{C}_\bfk^{-2} = { 2 \left(v_{N}^2\bfk^{2N} + D (D + \sqrt{D^2 + v_N^{2}\bfk^{2N}})\right)}$.
For what follows we will often use the notation $a_\bfk \equiv \mathcal{C}_\bfk \alpha_\bfk$ and $b_\bfk \equiv \mathcal{C}_\bfk \beta_\bfk$.
{Importantly, both the kinetic energy Eq. \eqref{eq_kinetic_energy} and the components of the wavefunction Eq. \eqref{eq_eigenbasis} transform well under the in-plane inversion symmetry $C_{2z}$.}
To characterize the nature of the Bloch state, it is helpful to consider the Berry curvature,
\begin{align}
    \Omega^{N} (\bfk) = -\frac{D N^2}{2} \frac{v_N^{2}\bfk^{2(N-1)}}{(v_N^{2}\bfk^{2N} + D^2)^{3/2} },
    \label{eq_berry_curv}
\end{align}
and the quantum metric,
\begin{align}
    \hspace{0mm} 
    \frac{g^N_{\mu \nu} (\bfk)}{\mathcal{G}_{\bfk}} = \begin{cases}
    D^2 + v_N^2 (\bfk^2 - k_\mu^2) \bfk^{2(N-1)}, & \mu=\nu \\    -v_N^2k_\mu k_\nu \bfk^{2(N-1)}, & \mu \neq \nu
    \end{cases}
    \label{eq_qm}
\end{align}
where $\{\mu,\nu\} = \{x,y\}$ and $\mathcal{G}_{\bfk}$ is defined in Appendix \ref{app_sec_quantumgeometry} along with additional details on the quantum geometric tensor.
The Berry curvature is strongly peaked about a ring located at $k_{\Omega}=
\left(\frac{2D^2}{v^2_N}\frac{N-1}{N+2}\right)^{\frac{1}{2N}}$ in momentum space (see inset of Fig.\ \ref{fig_band_berry_all} for R5G).
We coin this momentum-space ring where the Berry curvature peaks the `Berry ring of fire' (BRF).
For generic $N$-layers, the magnitude of the Berry flux enclosed by the ring is ${-}N\pi$.

\section{BCS pairing instability in multilayer graphene}
\label{sec_bcs_pairing_setup}

We examine superconducting instabilities triggered by attractive density-density interactions of the form
\begin{align}
    H_{\text{int}} = - \sum_{\bfr, \bfr'} V(|\bfr - \bfr'|) \rho(\bfr) \rho(\bfr').
\end{align}
Here the  $\rho(\bfr) = c_\alpha^{\dag}(\bfr) c_\alpha(\bfr)$, where $c_{\alpha} (\bfr)$ is the microscopic operator in the original orbital $\alpha = \{A_1, B_N\}$ basis; repeated indices are summed over.
The spin-valley index is dropped for simplicity as we are considering interactions of electrons of a given flavor ($K \uparrow$).
This attractive interaction is agnostic to the precise mechanism by which the electrons pair.
For the sake of analytic tractability, we consider an inversion-symmetric attraction, which in momentum space is given by, $V(\bfq) = U e^{-\alpha \bfq^2}$, where $\alpha$ and $U$ characterize the spatial-extension and the on-site strength of the interaction.
As will become important below, two asymptotic limits of the attraction length to consider are: (i) On-site local attraction, $\alpha \rightarrow 0$, and (ii) Long-range non-local attraction $\alpha \rightarrow \infty$.
The non-local limit, though seemingly artificial, provides an analytically-controllable limit to bring the quantum geometric quantities to the forefront.

Fourier transforming the above interaction in the BCS channel, and projecting the low-lying conduction electrons from the microscopic basis into the eigenbasis of Eq.\ \eqref{eq_eigenbasis}, i.e., $c_\alpha(\bfk) = u_{\alpha}(\bfk) \psi(\bfk)$, where $\psi(\bfk)$ is the band/eigen-basis, leads to a BCS-interaction of the form,
\begin{align}
    H_{\text{BCS}} 
    = - \sum_{\bfk, \bfk'} V_T(\bfk', \bfk) \psi^{\dag}(\bfk') \psi^\dag(-\bfk') \psi(-\bfk) \psi(\bfk),
\end{align}
where $V_T(\bfk', \bfk) = V(\bfk - \bfk') \langle u(\bfk') | u (\bfk) \rangle \langle u(-\bfk') | u(-\bfk) \rangle$ incorporates the form factors.
Since the pairing is between equal spin-valley electrons (where under parity $\psi(-\bfk) \psi(\bfk) = - \psi(\bfk) \psi(-\bfk)$ from fermionic anticommutation rules), the even-parity terms must be projected out of the pairing interaction to yield the correctly parity symmetrized pre-factor, 
\begin{align}
    V_T(\bfk', \bfk) &= V_A(\bfk', \bfk) \left[(a_{\bfk'}^*)^2(a_{\bfk})^2 + (b_{\bfk'}^*)^2(b_{\bfk})^2 \right] \\
    & \hspace{-2mm}+ 2 \left[V_S(\bfk', \bfk)\delta_{N, \mathbb{Z}_{\text{odd}}} + V_A(\bfk', \bfk)\delta_{N, \mathbb{Z}_{\text{even}}} \right] a_{\bfk'}^*a_{\bfk}b_{\bfk'}^*b_{\bfk},  \nonumber
\end{align}
where the parity-symmetric and -antisymmetric potentials are $V_{\frac{S}{A}}(\bfk', \bfk) = \big[V(\bfk' - \bfk) \pm V(\bfk' + \bfk) \big]/2$. Employing the aforementioned Gaussian form for $V(\bfq)$, the total interaction pre-factor can be written as,
\begin{widetext}
    \begin{align}
    V_T(\bfk', \bfk) = U e^{-\alpha (\bfk^2 + \bfk'^2)} \mathcal{C}_{\bfk}^2 \mathcal{C}_{\bfk'}^2 \Big[ &  v_N^4 \sinh{(2 \alpha \bfk \cdot \bfk')} |\bfk|^{2N} |\bfk'|^{2N} e^{2iN(\theta_{\bfk'} - \theta_{\bfk})}
   + \sinh{(2 \alpha \bfk \cdot \bfk')} \beta_\bfk^2 \beta_{\bfk'}^2
       \label{eq_vt_matrix_alpha}
 \\
    & + 2 v_N^2\beta_\bfk \beta_{\bfk'} |\bfk|^N |\bfk'|^N e^{iN(\theta_{\bfk'} - \theta_{\bfk})} \left( \cosh{(2 \alpha \bfk \cdot \bfk')} \delta_{N, \mathbb{Z}_{\text{odd}}} + \sinh{(2 \alpha \bfk \cdot \bfk')} \delta_{N, \mathbb{Z}_{\text{even}}} \right)   \Big], \nonumber \end{align}
\end{widetext}
where we have used the angular notation for momentum $k_x + i k_y = |\bfk|e^{i \theta_{\bfk}}$, and the other symbols have already been defined.
Performing the usual Hubbard–Stratonovich transformation (as described in Appendix \ref{app_sec_BCS}), we arrive at the BCS gap equation,
\begin{align}
    \Delta_{\bfk'} = \frac{1}{2} \sum_{\bfk} \frac{\tanh{(\beta E_\bfk/2)}}{E_{\bfk}} V_{T} (\bfk', \bfk) \Delta_{\bfk},
    \label{eq_gap_eqn}
\end{align}
where $\beta$ is the inverse temperature, $E_{\bfk} = \sqrt{\xi_{\bfk}^2 + |\Delta_\bfk|^2}$ and $\xi_{\bfk} = \epsilon (\bfk) - \mu$.
The chemical potential ($\mu$) is set for an electron density (above charge neutrality) of $10^{12}$cm$^{-2}$, which is the typical density range over which chiral superconductivity is observed in R4G \cite{han2025signatureschiralsuperconductivityrhombohedral}. Indeed, our findings are applicable for a range of electronic densities where the Fermi surface lies inside/outside the BRF.

The quarter-metal normal state (i.e., spin-valley polarized) dictates that the free energy is constrained by the symmetry group  
$G = [U(1)]_{S_z} \times [U(1)]_{V_z} \times [U(1)]_{\Phi} {\times [U(1)]_L}$, indicating the respective $U(1)$ symmetries of the $\hat{z}$-quantized spin- and valley-polarized flavors, the $U(1)$ gauge transformation of the superconducting phase, {and the $SO(2) \cong U(1)$ symmetry of the two-dimensional orbital momentum (where we anticipate a chiral gap function, as seen in the next sections).}
In the superconducting state,
the symmetry is reduced to the residual symmetry group $H = [U(1)]_{S_z-\Phi} \times [U(1)]_{V_z-\Phi} {\times [U(1)]_{L-\Phi}}$ where the phase accumulated under the combined operation of the spin, valley, phase, {and orbital momentum} rotations are compensated.
The superconducting order parameter transforms non-trivially under the symmetry 
group $R = G/H = [U(1)]_{S_z, V_z, \Phi, {L}}$ where there is a combined gauge-spin-valley-{orbital momentum} symmetry operation on all {four} spin, valley, phase, {orbital momentum} spaces.
The physical consequence of this symmetry classification is that the superconducting phase (or vortices) can wind an integer\footnote{This is unlike spontaneously spin-polarized spin-triplet superconductors \cite{berg2021_spin_triplet} and certain class of phases of superfluid $^{3}$He \cite{superfluid_spin_triplet_1986}, where two vortices annihilate each other and $4\pi$ phase windings unravel into a uniform configuration  i.e., $\pi_1 \left(SO(3)\right) = \mathbb{Z}_2$.} number of times i.e., $\pi_1\left(U(1)\right) = \mathbb{Z}$.

{It is important to note that our minimal model defined in Eq.\ \eqref{e1} misses the trigonal warping effect that is allowed by physical symmetries of R$N$G and causes significant distortion of Fermi surfaces \cite{macdonald_trilayer_2010}. Such effects lower the symmetry from $SO(2)$ down to $C_{3z}$ with a potentially important consequence for pairing: the kinetic energy no longer obeys $\epsilon(\bfk)=\epsilon(-\bfk)$ which tends to suppress the susceptibility for the formation of $(+\bfk,-\bfk)$ Cooper pairs. In this situation, pairs (from a given valley) may instead form at a non-zero center-of-mass momentum \cite{christos2025finite}, although the critical temperature of such a state typically tends to be significantly lower. By contrast experiments in R4G and R5G show $T_c$ as high as 300 mK which is not small given the very low carrier density in these materials. We take this to indicate prevalence of conventional $(+\bfk,-\bfk)$ pairing, which should be well described by the minimal model based on Hamiltonian $H_0$ in Eq.\ \eqref{e1}.}

\section{Local on-site attraction: $\alpha = 0$}
\label{sec_local_sol}

\begin{figure}[t]
\includegraphics[width = 0.47\textwidth]{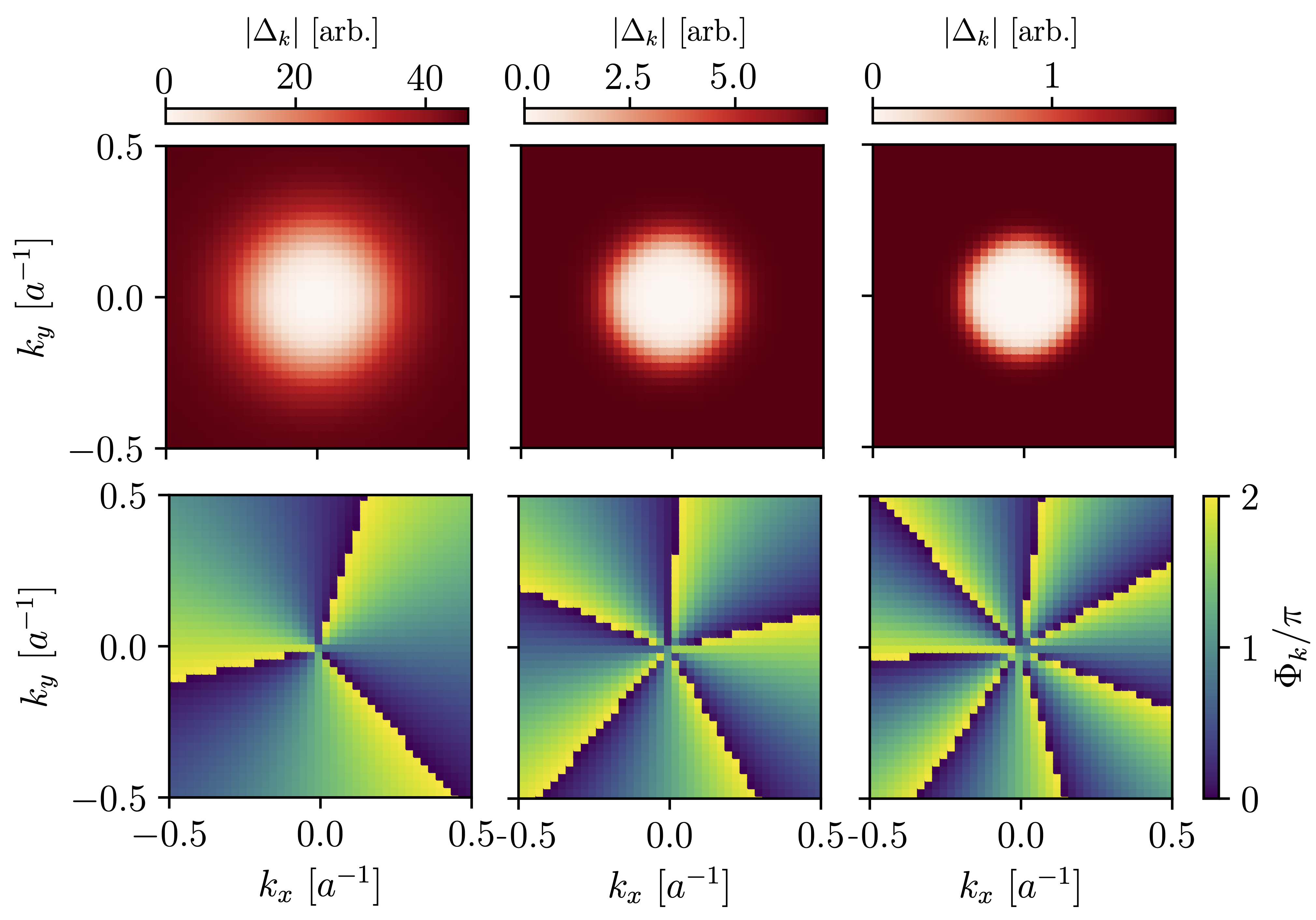}
\caption{Superconducting order parameter magnitude (top) and phase (bottom) for [left to right] $N=3,5,7$ layers for on-site local attraction ($\alpha = 0$), and $U = 10^{-6} a$ eV/m.} 
\label{fig_sc_odd_alpha_0}
\end{figure}

We first consider the {special} limit of local (in real-space) interactions, corresponding to $\alpha=0$.
From inspection of the interaction potential in Eq.\ \eqref{eq_vt_matrix_alpha}, we note the sole surviving term is $\cosh(...) \delta_{N, \mathbb{Z}_{\text{odd}}}$, which explicitly depends on the non-trivial Bloch form factor.
This indicates that local pairing can trigger -- for same flavor fermions -- pairing instabilities only for odd-$N$, and not for even-$N$.
This arises from the preservation of Fermi-Dirac statistics (and spatial parity symmetry) for odd-parity Cooper pairs.
Figure \ref{fig_sc_odd_alpha_0} presents the magnitude and phase of the superconducting gap for representative odd $N$-layer systems ($N = 3,5,7$) for on-site ($\alpha=0$) local interactions. 
As seen, the superconducting order parameter winds $N$-times for R$N$G, and spontaneously breaks the full rotational symmetry.
In this limit, the zero-temperature order parameter can be extracted from the gap equation from inspection,
\begin{align}
    \Delta_{\bfk}
    =  \Delta_0 \frac{|\bfk|^N }{\sqrt{\bfk^{2N} + D^2}} e^{i \theta_\bfk N},
    \label{eq_alpha_0_ansatz}
\end{align}
where $\Delta_0$ is determined by solving the Eq.\ \eqref{eq_gap_eqn} with the ansatz of Eq.\ \eqref{eq_alpha_0_ansatz}.
The behavior of the gap is such that a node exists at $\bfk = 0$ (arising from the vanishing Bloch wavefunction component $\alpha_\bfk$) and the magnitude of the gap grows monotonically to its asymptotic value of $\Delta_0$.
The phase winding of $N$ times reflects a $f, h, j,...$ angular momentum channel for $N = 3, 5, 7, ...$ layers.
A range of chiral (broken time-reversal and broken spatial inversion) superconductors of the form {$(p_x+ip_y)^N$} are thus possible for local attractions between electrons in odd-$N$ layered systems.
{
The chirality of the gap directly follows from the $a_{\bfk}$ component in Eq.~\eqref{eq_eigenbasis}, thus emphasizing the role of the electronic Bloch wavefunction in determining the pairing gap.
We note that this striking conclusion of chiral superconductivity arising only in odd-$N$ layers is special to the contact-limit of the attraction and to systems where $C_{2z}$ is a symmetry.}

\section{Extended attractive interactions}
\label{sec_extended_int}

\begin{figure}[t]
\includegraphics[width = 0.47\textwidth]{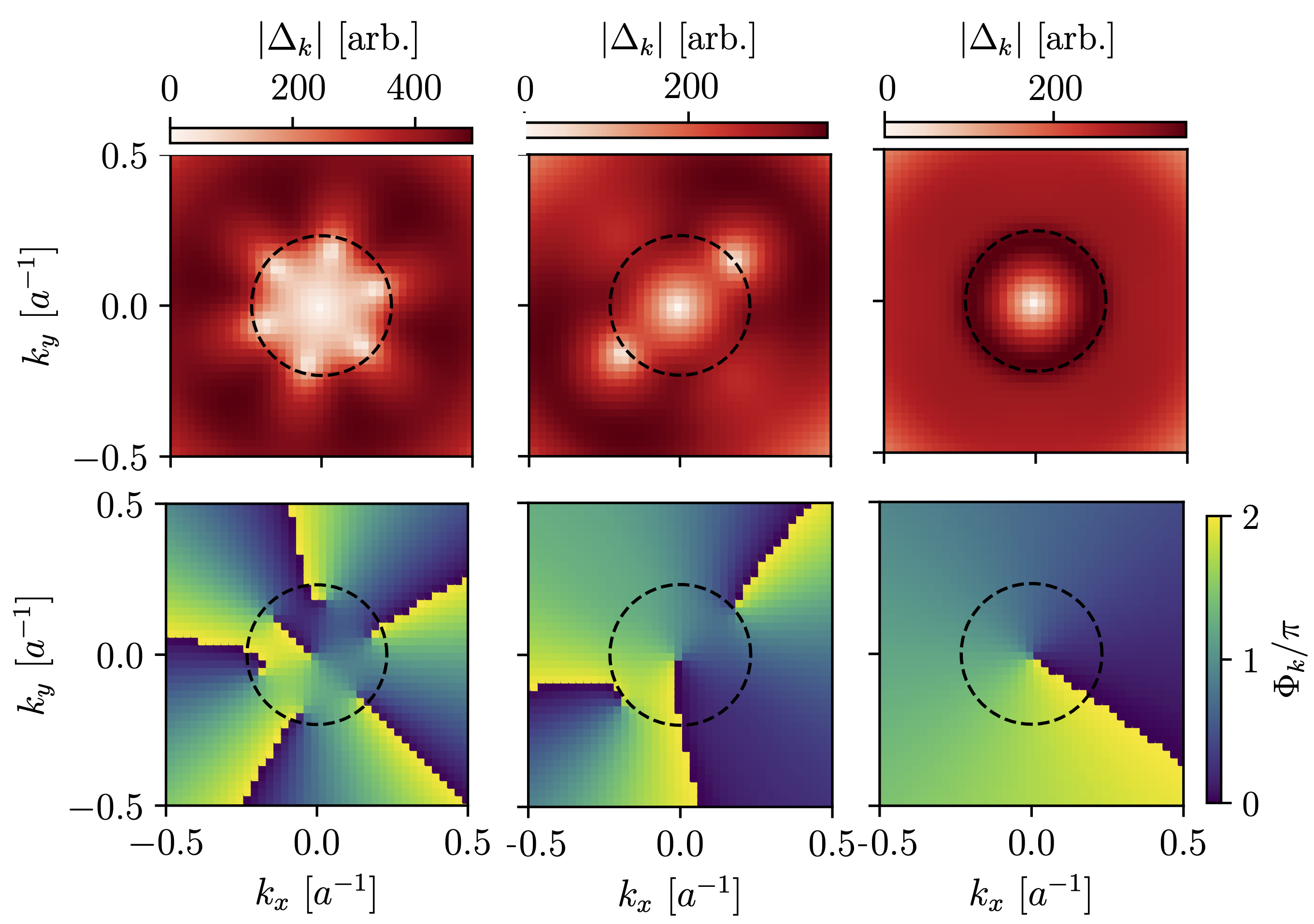}
\caption{Superconducting order parameter solutions magnitude (top) and phase (bottom) for odd-$N$ layers ($N=5$) for non-local attraction ($\alpha = 20.4 a^{2}$).  
The black dashed circle indicates where Berry curvature is peaked. $U = 10^{-4} a$ eV/m.}
\label{fig_sc_r5g_alpha_fixed}
\end{figure}

In the presence of spatially-extended interaction ($\alpha \neq 0$), additional pairing channels can be realized.
Figure \ref{fig_sc_r5g_alpha_fixed} presents a subset of possible pairing channels realized for R5G for attraction length $\alpha = 20.4 a^{-2}$.
Superimposed on the distribution of the superconducting order parameter is the afore-described `BRF' (black dashed line), where the Berry curvature peaks, at a radius of $k_{\Omega}$.
For $|\bfk|<k_{\Omega}$, there is a single winding of ${\pm}2\pi$ of the phase of the superconducting order parameter.
For $|\bfk| > k_{\Omega}$ and well outside the BRF, R5G possesses chiral gap solutions of winding (i) $N=5$, (ii) $N-2=3$, and (iii) single-winding (in units of $2\pi$).
The single-winding solution is the familiar $p_x\pm ip_y$ superconducting order parameter, which is the lowest harmonic odd-parity gap that benefits from the condensation energy gain; this is a natural solution that one obtains from a uniform in momentum space attractive potential.
We present in Appendix \ref{app_r7g_vortex_nucleation} two instances of the single-winding phase solution for R7G, where the phase-winding inside the BRF is of the same/opposite chirality to the phase-winding outside the BRF.
We emphasize that these -- and additional odd-parity chiral -- solutions are independent of the displacement field strength (see Fig.\ \ref{fig_r5g_smaller_d} in Appendix \ref{app_smaller_d_r5g} for similar gap solutions realizable at smaller $D$).

\begin{figure}[t]
\includegraphics[width = 0.47\textwidth]{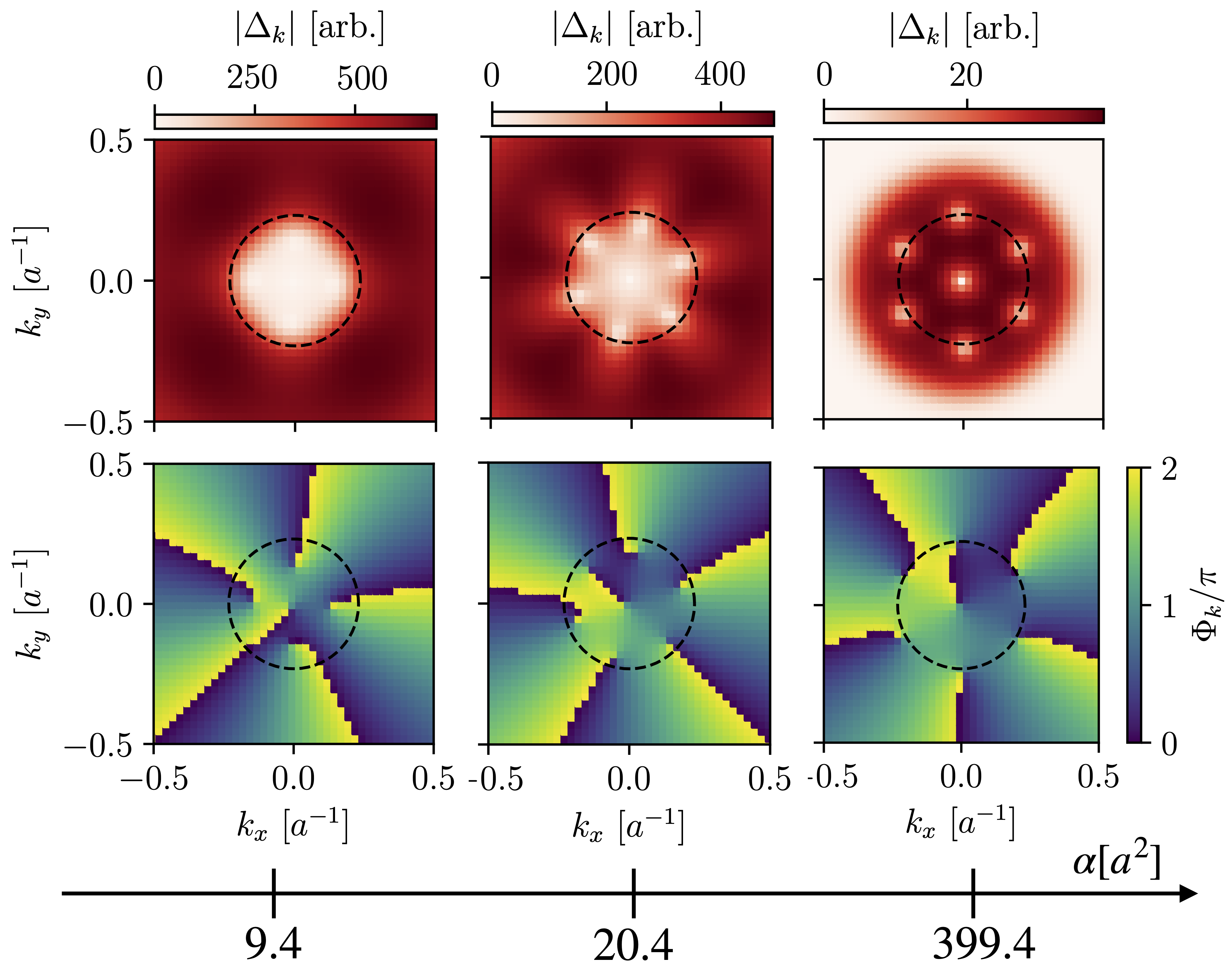}
\caption{Evolution of magnitude (top) and phase (bottom) of $N$-times phase winding order parameter for $N=5$ layers as a function of attraction length $\alpha$. 
The black dashed circle indicates where the Berry curvature is peaked. $U = 10^{-4} a$ eV/m. Figures \ref{fig_evolution_alpha_N_5_Nm2_winding} and \ref{fig_evolution_alpha_N_5_single_winding} in Appendix \ref{app_r5g_evolution_alpha} depict the evolution of the $N-2$ and single-winding solutions under increasing $\alpha$.} 
\label{fig_evolution_alpha_N_5_N_winding}
\end{figure}

For ever increasing length of attractions, as depicted in Fig.\ \ref{fig_evolution_alpha_N_5_N_winding}, a sharp change from the $p_x\pm ip_y$ order parameter behavior to higher phase-windings occurs increasingly closer to the BRF (for smaller $\alpha$, this change occurs for momenta just inside the ring, $|\bfk|\lesssim k_{\Omega}$).
Indeed, for asymptotically large $\alpha$, this change occurs precisely at select momentum locations \textit{on} the BRF: the order parameter magnitude is suppressed and the phase winds i.e., momentum-space vortices of the superconducting order parameter at equal spacing are nucleated on the BRF.

\begin{figure}[ht]
\includegraphics[width = 0.47\textwidth]{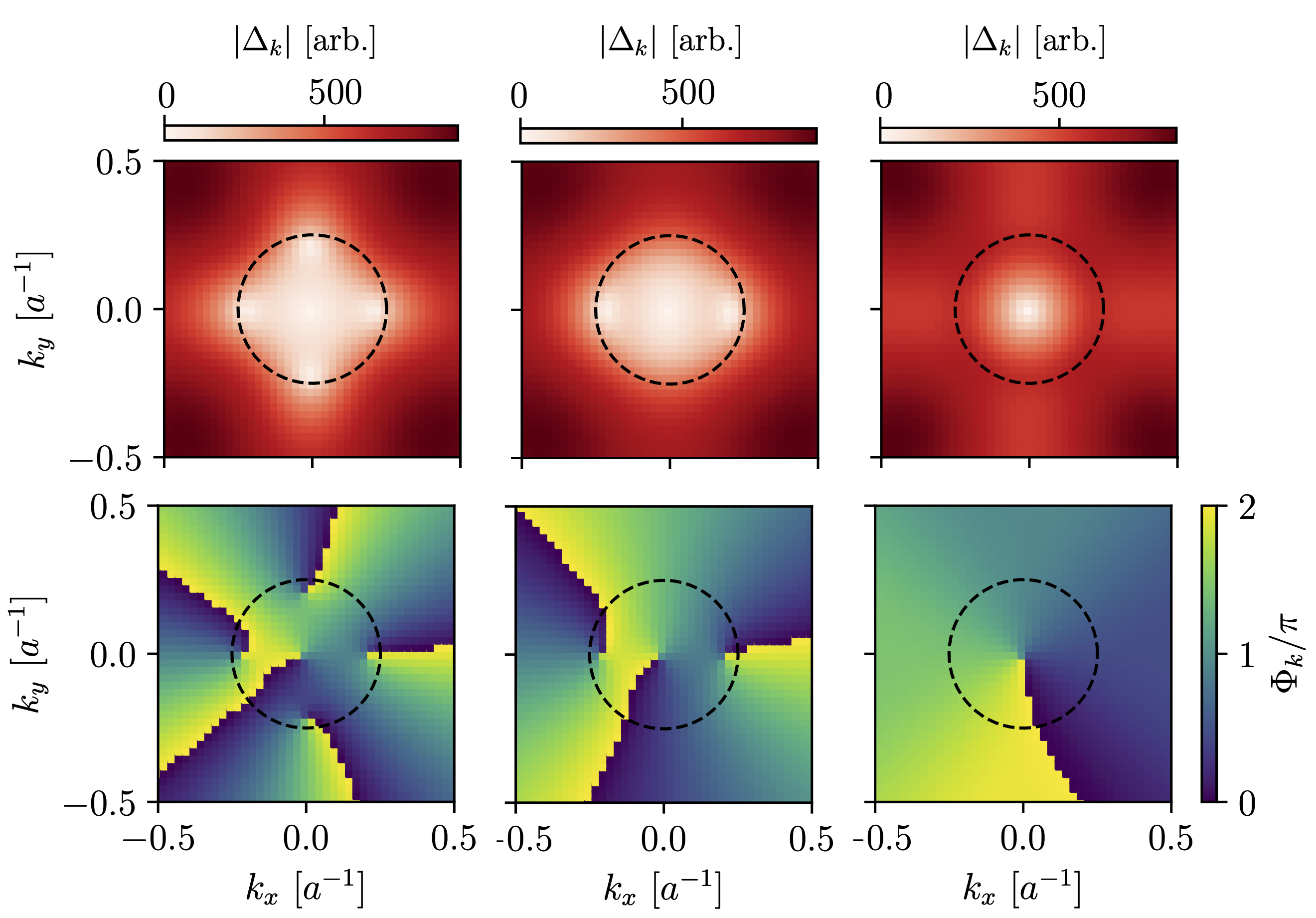}
\caption{Superconducting order parameter solutions magnitude (top) and phase (bottom) for even-$N$ layer (R4G) for non-local attraction ($\alpha = 9.4 a^{2}$).  
The black dashed circle indicates where Berry curvature is peaked. $U = 10^{-4} a$ eV/m.} 
\label{fig_sc_r4g_alpha_fixed}
\end{figure}

Turning to the even-$N$ structures, we present in Fig.\ \ref{fig_sc_r4g_alpha_fixed} the gap solutions for R4G for a fixed $\alpha$.
Similar to odd-$N$ layers, for $|\bfk|<k_{\Omega}$, there is a single-winding solution.
Outside the BRF, possible solutions include $N\pm1 = \{3,5\}$, as well as a continuation of the single-winding solution.
Only odd winding solutions are permissible regardless on if $N$ is odd or even, which is a consequence of accommodating odd-parity channels (due to Fermi-Dirac statistics), leading to a family of chiral superconductors.

The key point is that higher-winding odd-parity chiral solutions are found \textit{outside} the BRF for both odd and even $N$ layered systems.


\subsection{Extended in real space interaction: small $\alpha$ limit}
\label{sec_sub_alpha0}

The variety of pairing solutions can be unveiled from expanding the interaction potential Eq.\ \eqref{eq_vt_matrix_alpha} in the small-$\alpha$ limit, where to linear order in $\alpha$,
\begin{align}
    \hspace{0mm} V_T(\bfk', \bfk) & \approx 2 U  A_{\bfk', \bfk}^{N} e^{iN(\theta_{\bfk'} - \theta_{\bfk})}\delta_{N, \mathbb{Z}_{\text{odd}}}     \label{eq_small_alpha} \nonumber \\
  &+\alpha U  \cos(\theta_{\bfk} - \theta_{\bfk'}) \Big[  B_{\bfk', \bfk}^{N} 
+C_{\bfk, \bfk'}^{N}e^{2iN(\theta_{\bfk'} - \theta_{\bfk})} \nonumber 
\\ & \ \ \ \ \ +D_{\bfk', \bfk}^{N} e^{iN(\theta_{\bfk'} - \theta_{\bfk})} \delta_{N, \mathbb{Z}_{\text{even}}} \Big] ,
\end{align} 
where $A_{\bfk, \bfk'}^{N}$ etc.\ are $\theta$-independent factors given in Appendix \ref{app_higher_order_expansion}. 
Crucially, the bare interaction introduces additional phase winding for the order parameter.
The term in the first line contains the already examined local interaction term with a winding-less renormalization at linear order in $\alpha$, while the remaining terms promote alternate pairing channels, in the order in which they appear: (i) single-winding, (ii) $2N \pm 1$ winding, and (iii) $N\pm 1$ winding for even $N$.
In Appendix \ref{app_higher_order_expansion} we generalize the results to higher powers in small-$\alpha$, where one finds that for odd-$N$ phase winding of $N \pm 2 \mathbb{Z}$ is also possible.

Importantly, all of the interaction matrix elements in Eq.\ \eqref{eq_small_alpha} are weighted by momentum dependent form factors.
As such, for small momenta, the least-suppressed momentum-dependent matrix element is the one which produces the $(p_x\pm ip_y)$ gap for the even-$N$ layers; for odd-$N$ layers, this can even triumph energetically over the $\alpha=0$ matrix element that gives rise to $N$-times winding but is weighted by momentum $|\bfk|^N$.
At larger momenta, the higher phase-winding pairing channels can be energetically preferred over the single phase-winding (as their corresponding matrix element is weighted in higher powers of $|\bfk|$).
Intriguingly, the momentum-space location about which the different pairing channels surface is the BRF, even though the formalism encapsulates the entire wavefunction and not just the local quantum geometry.

\subsection{Extended in real space interaction: large $\alpha$ limit}
\label{sec_sub_alphainf}

The BRF is evidently playing a crucial role in the change of the phase winding.
To extract out the origin of this phase-change, we examine the other asymptotic limit, namely taking $\alpha \rightarrow \infty$.
Corrections about this asymptotic limit introduces a momentum space `spread' from the overlap of the Bloch functions at infinitesimal momentum separation ($|\bfq| = |\bfk' - \bfk| \ll 1/ \sqrt{\alpha}$),
\begin{align}
    \langle u(\bfk') | u (\bfk) \rangle &= \langle u(\bfk + \bfq) | u (\bfk) \rangle  \\
    &\approx \exp\left[{-\frac{g_{\mu \nu}(\bfk)}{2} q_\mu q_\nu}\right] \exp\left[ -i \int_\bfk ^{\bfk + \bfq} d \bfl \cdot \vec{\cala}(\bfl) \right] \nonumber ,
\end{align}
where $\vec{\mathcal{A}}(\bfl)$ and $g_{\mu \nu}(\bfk)$ are the Berry connection and quantum metric.
In this limit the BCS-interaction\footnote{The interaction is gauge invariant, despite the appearance of the Berry connection, as the phases accumulated under the gauge transformations $\cala_\mu(\bfk) \rightarrow \cala_\mu(\bfk) + \partial_\mu \theta(\bfk)$ and $\ket{ u(\bfk) } \rightarrow \ket{u(\bfk)} e^{i \theta(\bfk)}$ cancel.} is, 
\vspace{-10mm}
\begin{widetext}
\begin{align}
   \hspace{-8mm} H_{\text{BCS}} &= -U \sum_{\bfk, \bfq} \exp\left[-{(\alpha \delta_{\mu \nu} + g_{\mu \nu}(\bfk)) q_\mu q_\nu}\right] \exp\left[ -i \int_\bfk ^{\bfk + \bfq} d \bfl \cdot \vec{\cala}_-(\bfl) \right] \psi^{\dag}(\bfk+\bfq)\psi^{\dag}(-\bfk-\bfq) \psi(-\bfk)\psi(\bfk),     \label{eq_bcs_int_qg}
\end{align}
\\
\end{widetext}
where $\vec{\cala}_-(\bfl) \equiv \vec{\cala}(\bfl)- \vec{\cala}(-\bfl)$ and we have taken the quantum metric to be an even function in momentum, as is the case in Eq.\ \eqref{eq_qm}.
This quantum geometric dependent pairing potential, just as was done in Sec.\ \ref{sec_bcs_pairing_setup}, must be properly anti-symmetrized to enforce the odd-parity nature of the pairing channel
\begin{widetext}
    \begin{align}
    \hspace{-6mm} V_{\rm QG} (\bfk_2, \bfk_1) &=  \frac{U}{4}\sum_{\eta_{1,2} = \pm1} \eta_1 \eta_2 
    \exp{ \Big[ -\left(\alpha \delta_{\mu \nu} + g_{\mu \nu}(\bfk_1)\right) (\eta_2 \bfk_2 - \eta_1 \bfk_1)_\mu(\eta_2 \bfk_2 - \eta_1 \bfk_1)_\nu \Big]}  \exp{\left[ -i \int_{\eta_1 \bfk_1} ^{\eta_2 \bfk_2} d \bfl \cdot \vec{\cala}_-(\bfl) \right]}. 
\end{align}
\end{widetext}
The BCS-Hamiltonian (and the associated free energy by taking the expectation value) resembles that of a superconducting order parameter in a momentum-space magnetic field -- with the Berry connection taking the role of the magnetic vector potential -- and the pairing interaction strength suppressed by the quantum metric.
This observation can be seen more explicitly from the saddle-point gap equation,
\begin{widetext}
\begin{align}
    \hspace{-2mm} 
    |\Delta_{\bfk_2}| = \frac{U}{4} & \sum_{\bfk_1} \sum_{\eta_{1,2} = \pm1}  \eta_{1} \eta_2 \exp{ \Big[ -\left(\alpha \delta_{\mu \nu} + g_{\mu \nu}(\bfk_1) \right) (\eta_2 \bfk_2 - \eta_1 \bfk_1)_\mu(\eta_2 \bfk_2 - \eta_1 \bfk_1)_\nu \Big]} \label{eq_gap_qg} \\
    &\times \exp{\left[ -i \int_{\eta_1 \bfk_1} ^{\eta_2 \bfk_2} d \bfl \cdot \left(\vec{\cala}_-(\bfl) + \del_\bfl\Phi \right) + \left(\frac{4n+1-\eta_1 \eta_2}{2}\right)i\pi \right] } \tanh\left({\beta E_{\bfk_1}\over 2}\right)\frac{|\Delta_{\bfk_1}|}{2 E_{\bfk_1}} \nonumber ,
\end{align}
\end{widetext}
where the order parameter is expressed as $\Delta_\bfk = |\Delta_\bfk| e^{i \Phi_\bfk}$ and $n \in \mathbb{Z}$.
The additional phase of $2n\pi$ or $(2n+1)\pi$ is due to the odd spatial parity i.e., $\Delta_{-\bfk} = -\Delta_{\bfk}$.
Ensuring that the right-hand side of the gap equation is a real quantity (to complement the left-hand side) for each $\eta_{1,2}$ leads to a flux condition of,
\begin{align}
\hspace{-3mm} \Delta  \Phi|_C = & - \int_{{C}} d \bfl \cdot  \left(\vec{\cala}(\bfl)- \vec{\cala}(-\bfl) \right) + \pi n, 
\label{eq_flux_quantization}
\end{align}
where $\Delta \Phi|_C$ is the phase accumulated under the path ${C}$, and the second term on the right-hand side specifies whether an even or odd multiple of $\pi$ is accumulated depending on the path taken.
Over a closed loop about the origin, we thus arrive at the winding condition of the superconducting phase,
\begin{align}
    \Delta \Phi_{\partial \Sigma} & = - 2\oint_{\Sigma}  \vec{\Omega} \cdot d{\mathbf{\Sigma}} + 2\pi n_{\rm odd/even} \label{eq_closed_loop_flux_quantization_stokes}
    \\
     & = - 2\Phi_{\Omega} + 2\pi n_{\rm odd/even},
\label{eq_closed_loop_flux_quantization}
\end{align}
where $n_{\rm odd/even}$ is an odd/even integer, $\partial \Sigma$ is the boundary of the closed loop that encloses the disk $\Sigma$.
We note that we have exploited the odd symmetry of the Berry connection in our gauge choice i.e., $\vec{\cala}(-\bfl) = -\vec{\cala}(\bfl) + \nabla_\bfl\chi$.
We direct the reader to Appendix \ref{app_sec_quantumgeometry} where we discuss how the even-in-momentum nature of the Berry curvature (as is the case in Eq.\ \eqref{eq_berry_curv}) entails at least one gauge choice where the Berry connection is an odd function in momentum; any other terms arising from the gauge transformation ($\nabla_\bfl\chi$) would  of course vanish under application of Stokes' theorem in Eq.\ \eqref{eq_closed_loop_flux_quantization_stokes}.

Importantly, it is the single-valued nature of the complex order parameter that implies $\Delta \Phi_{\partial \Sigma} = 2\pi n$, where here $n$ must be an odd integer so that the gap function obeys the appropriate statistics and spatial parity symmetry. 
From Eq.\ \eqref{eq_closed_loop_flux_quantization} the phase winding is thus composed of two pieces.
The first-term is merely the momentum-space equivalent of the flux-quantization condition for a superconducting ring in a pierced background magnetic field, with the Berry curvature taking the role of the external magnetic field \cite{tinkham2004introduction}.
The second term is what we call the `statistical flux', which for a closed loop contributes the necessary phase to ensure that the overall winding of the superconducting phase is an odd-multiple of $2 \pi$ in accordance with Fermi-Dirac statistics of spin-valley triplet superconductors.

As such, the winding of the superconducting order parameter is determined by the sum of twice the pierced/enclosed Berry flux\footnote{In the context of a crystal order parameter, there is an alternative flux-quantization condition that determines the Chern number of a Hartree-Fock band \cite{dong2024stability}.} (rounded to the nearest integer multiple of $\pi$), and even/odd integer multiples of $2\pi$ to ensure an overall odd parity to the gap function.
This indicates that the Berry flux can be regarded as the `anchor-point' about which superconducting phase winding can be determined, but is not the sole determining factor in the overall phase winding.

Taking odd-$N$ layers, from the above flux quantization condition, for enclosed Berry-flux of $N \pi$ (i.e., enclosing the entire BRF), one thus correctly presumes a gap solution of $N$-times (in units of $2\pi$) phase-winding is possible.
This can be accommodated, for example, by the nucleation of $N-1$ vortices (left-panel of Fig.\ \ref{fig_evolution_alpha_N_5_N_winding}) or by $N+1$ vortices (right-panel of Fig.\ \ref{fig_evolution_alpha_N_5_N_winding}) on the BRF depending on if the additional vortex at the origin has a $+1$ or $-1$ relative chirality.
In this sense, the Berry curvature is playing a similar role as the magnetic field -- with the number of vortices growing with the magnitude of the Berry flux enclosed by the BRF -- analogous to the number of nucleated real-space vortices in a type-II superconductor being proportional to the strength of the magnetic field.
{It is interesting to note the special $\alpha = 0$ limit (middle panel of Fig.~\ref{fig_sc_odd_alpha_0}) has a phase winding of $N$-times, which is the same as a mean field solution of $N$-times phase winding at finite $\alpha \neq 0$ (Fig.~\ref{fig_evolution_alpha_N_5_N_winding}) for $|\bfk| > k_{\Omega}$. 
This evident smooth connection as $\alpha$ increases for this mean-field solution is a consequence of analytic continuity of the model we study. As such, though the momentum-space flux quantization is formally derived in the asymptotic $\alpha \rightarrow \infty$ limit, the change in the phase winding and nucleation of vortices are expected (and do appear) even at a finite spatial length for the attraction.
}

The important subtlety in this analogy is the statistical flux, which can give rise to additional phase-windings to comply with the overall odd-parity nature of the superconductor.
This was evident in our numerical gap solutions of $N-2$ winding accommodated by $N-3$ vortices (middle panel of Fig.\ \ref{fig_sc_r5g_alpha_fixed}) for R5G; similarly, for even-$N$ systems (Fig.\ \ref{fig_sc_r4g_alpha_fixed} for R4G), the enclosed Berry flux is an even multiple of $2\pi$, and as such the statistical flux is required to generate the odd-parity winding solutions outside the BRF.
Intriguingly, there are also single-winding gap solutions (right-panel of Fig.\ \ref{fig_sc_r5g_alpha_fixed} for R5G; also see Fig.\ \ref{fig_r7g_single_winding_vortices_or_not} in Appendix \ref{app_r7g_vortex_nucleation} for R7G) that are apathetic or cognizant to the presence of the Berry flux leading to the nucleation of zero or a single vortex, respectively.
In other words, the magnitude of the enclosed Berry curvature ($N\pi$) does not necessarily translate into the number of nucleated equally-spaced, momentum-space vortices for the mean-field solutions.
What is clear, nonetheless, is that the Berry curvature facilitates these additional gap solutions by nucleating vortices; but due to the phase-winding also being set by the statistical flux and the presence of additional (not associated with the Berry flux) momentum space vortices, this momentum-space field does not have a uniquely identical effect as a real-space magnetic field on the superconducting order parameter. 

The sign in front of the Berry flux term in Eq.\ \eqref{eq_flux_quantization} may give the impression that an `anti-flux quantization' condition (and a subsequent anti-Little-Parks effect) is being determined by the Berry curvature.
This, however, is merely an artifact of the sign convention used in our definition of the Berry connection (see Appendix \ref{app_sec_quantumgeometry}); by flipping the sign-convention of the Berry connection, we would be led to the standard flux quantization and Little-Parks effect \cite{little_parks_1962}.
Recently, Ref. \cite{may2025pairing} has interestingly proposed that short-range attractive potentials and overscreened repulsive potentials in the presence of Berry curvature exhibit `opposite-sign' flux quantization conditions, respectively.
We note that the result for the overscreened potential utilized the eigenfunction form-factors {of the full $2N$-orbital model} within the diagrammatic construction of the potential; a mere wavefunction projection {of our $k^N$ minimal model} on top of a standard overscreened potential is not sufficient to generate this `opposite-sign' flux quantization condition.

A further non-trivial finding is that though this magnetic-field analogy is established formally in the large $\alpha$ limit, nonetheless the effects are also noticeable for smaller $\alpha$ (as seen in Fig.\ \ref{fig_sc_r5g_alpha_fixed} and \ref{fig_sc_r4g_alpha_fixed}).
The reason for this is that the positive-definite quantum metric plays a similar role as the attraction length $\alpha$ by weighting the interaction -- albeit by a momentum-space and non-isotropic attraction -- length to yield a renormalized `effective' attraction length 
$\alpha_{\mu \nu}^{\text{eff}}(\bfk) = \alpha \delta_{\mu \nu} + g_{\mu \nu}(\bfk)$.
This enlargement of the spatial extension of the attraction by $g^N_{\mu \nu}(\bfk)$ is in agreement with the notion of the quantum metric providing a measure of the spatial-spread of the electronic wavefunction in real space \cite{marzari1997maximally}.
The upshot is that the real and imaginary components of the quantum geometric tensor are playing two roles: (i) the Berry curvature, as a momentum-space magnetic field, can introduce vortices to suppress the superconducting order parameter, and (ii) the quantum metric extends the spatial range of an otherwise nominally local interaction.

With this flux condition addressed, the suppression of the order parameter at these momentum-space vortices can be seen more explicitly by retaining the leading term in the small $\bfq$ expansion of the order parameter $\Delta_{\bfk + \bfq}$, and extending the momentum summation to $\pm \infty$ in Eq.\ \eqref{eq_gap_qg} (the exponentially decaying interaction terms makes this approximation reasonable) to yield,
\begin{align}
  \hspace{-1.9mm}  |\Delta_\bfk|^2 
    &=  {\left(\frac{U}{4 \pi}\right)^2\frac{1}{\alpha^2 + \alpha \text{tr}[g^N(\bfk)] + [\Omega^N(\bfk)]^2/4 } - \xi_{\bfk}^2} 
    \label{eq_gap_suppressed_full} \\
    &\hspace{-1.9mm}  \approx \Big[ {\left(\frac{U}{4 \pi \alpha}\right)^2  - \xi_{\bfk}^2 \Big] - \left(\frac{U}{4 \pi \alpha}\right)^2 \frac{\text{tr}[g^N(\bfk)]}{\alpha}} + ... 
\end{align}
where in the final line we have taken the large $\alpha$ expansion; {see Appendix~\ref{app_sec_qg_gap_influence} for details of the derivation.}
We notice that the magnitude of the order parameter is suppressed where the trace of the quantum metric is peaked.
Since, $\text{tr}[g^N(\bfk)] = \frac{|\Omega^N (\bfk)|}{2} \frac{2 D^2 + \bfk^{2N}}{D \sqrt{D^2 + \bfk^{2N}}}$, where the momentum dependent form factor is a monotonically increasing function in $\bfk$, the peak in the trace quantum metric corresponds to the peak in the Berry curvature.
This suppression of the order parameter by the magnetic field is physically consistent with our established Berry curvature and magnetic field analogy.

\section{Density and field tuned transition between chiral superconductors}
\label{sec_transition_chiral_sc}
\begin{figure}[t]
\includegraphics[width = 0.48\textwidth]{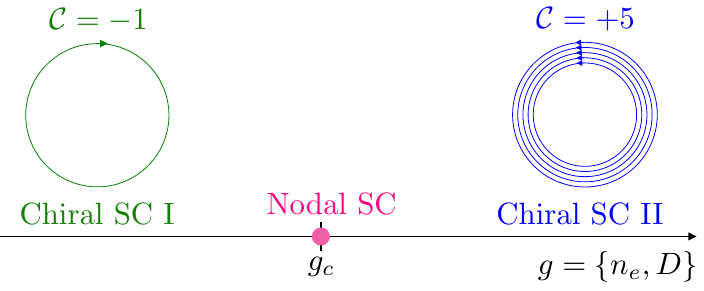}
\caption{Transition between two distinct chiral superconductors (`Chiral SC I, II' with Chern number $\mathcal{C}=-1,5$) as a function of electronic density ($n_e$) or displacement field ($D$). A gapless nodal superconductor (`Nodal SC') separates the two chiral superconductors.}
\label{fig_chiral_sc_transition}
\end{figure}
The nucleated vortices on the BRF suggests that the system can be driven into and out of its chiral superconducting state by tuning the magnitude of the displacement field or the electronic density.
Let us consider the example, where a vortex of chirality $-1$ is situated at the origin, and the BRF contains six vortices of chirality $+1$ each, as in the right-panel of Fig.\ \ref{fig_evolution_alpha_N_5_N_winding}.
For a generic Fermi surface that encircles the origin, but is smaller than the radius of the BRF, the $\textbf{d}(\bfk)$ vector for the associated Bogoliubov-de-Gennes (BdG) Hamiltonian winds once about the origin, yielding a Chern number of $\mathcal{C} = - 1$ with the sign determined by the chirality of the {$p_x - ip_y$ gap situated at the origin}.
Increasing the electron density (while keeping the displacement field fixed) -- or alternatively, decreasing the displacement field while keeping the electron density fixed -- shrinks the BRF relative to the Fermi surface leading to the inclusion of the nucleated vortices inside the Fermi surface.
The winding of the associated $\textbf{d}(\bfk)$ vector is now clearly influenced by the chirality of all the vortices i.e., the vortex at the origin as well as the nucleated vortices on the BRF such that $\mathcal{C} = 6 - 1 = 5$, where six ($+1$) vortices are generated on the ring, one of which terminates the winding from the interior $p_x-ip_y$ gap ($-1$) vortex. 

Figure \ref{fig_chiral_sc_transition} depicts such an electron density ($n_e$) or displacement field ($D$) tuned transition, with the tuning parameter $g = \{n_e, D\}$.
The two gapped chiral superconductors are labeled as `Chiral SC I/II' which have a respective BdG Chern number of $-1$ and $+5$ and associated number of edge states for a finite-sized system.
A quantum phase transition separates these two chiral superconductors at $g_c$, where the Fermi surface overlays the BRF, to yield a gapless nodal superconductor, with the number of nodes dictated by the number of vortices.

\section{Experimental Signatures}
\label{sec_expt_signatures}

Nearly all of the superconductors predicted in this work are topologically non-trivial ({the gapped chiral superconductors belong to the class D in the ten-fold way classification \cite{ryu2010topological,hasan2010colloquium}}) and will thus exhibit protected chiral gapless edge modes. In principle at least, signatures of such edge modes can be detected either directly by scanning tunneling spectroscopy \cite{Madhavan2021} or by probing magnetic fields induced by the edge currents \cite{Kallin2016} through scanning magnetometry techniques. In practice this type of detection will be challenging, due to the necessity of dual gating that is required to tune R$N$G into the superconducting state but at the same time obstructs easy access to sample surfaces. Samples will also exhibit a quantized thermal Hall effect, which is however notoriously difficult to measure.    

The transition between two different chiral phases, however, could be detected by a bulk measurement. Specifically, we envision probing the nature of bulk quasiparticle excitations using the newly developed {kinetic inductance} technique  \cite{banerjee2025superfluid,tanaka2024superfluidstiffnessflatbandsuperconductivity} that enabled highly accurate measurements of the temperature-dependent superfluid stiffness $\rho_s(T)$ in twisted bilayer and trilayer graphene. In fully gapped SC phases one expects an exponentially activated behavior $\delta\rho_s(T) =\rho_s(T)-\rho_s(0)\sim e^{-\Delta/k_B T}$ at low $T$, where $\Delta$ denotes the excitation gap. 
Exactly at the transition gapless quasiparticles will give rise to a power law behavior. From the extensive work on $d$-wave superconductors in the context of the high-$T_c$ cuprates, it is known that point-nodes in two spatial dimensions give  $\delta\rho_s(T) \sim T$ in the clean limit, crossing over to $\sim T^2$ behavior in the presence of disorder \cite{Hirschfeld1993}. Hence, it should be possible to identify the critical point from this type of measurement. In fact all physical observables, including the specific heat $C_V(T)$, will exhibit such a power law/activated behavior when tuned through the critical point.   

In analogy with spinless $p_x\pm ip_y$ we expect topological phases with odd ${\cal C}$ to harbor Majorana zero modes in cores of (real-space)  Abrikosov vortices. Each pair of such vortices  can be pictured as encoding a single complex fermionic zero mode implying an extensive ground state degeneracy $2^{M/2}$ in a system with $M$ vortices. The corresponding residual ground-state entropy $S_0$ can be measured in electronic nanostructures via the temperature dependence of charge transitions \cite{Sela2019}, relying on a Maxwell thermodynamic relation that relates changes in the entropy $S$ with chemical potential $\mu$ to changes in the particle number $N$ with temperature $T$. This technique could be applied to our system in the presence of transverse magnetic field $B$ which will nucleate $M=\Phi_B/\Phi_0$ Abrikosov vortices, where $\Phi_B$ is the total magnetic flux and $\Phi_0=hc/2e$ is the superconducting flux quantum.

\section{Discussion}
\label{sec_discussion}

Inspired by recent experimental findings of unconventional superconductivity in rhombohedral-stacked $N$-layer graphene (R$N$G) 
we examined, in this work, the influence of the Bloch wavefunction and quantum geometry on the nature of superconductivity within a minimal model of R$N$G. Specifically, we considered a single-band model of spin and valley polarized conduction electrons with a flat band-dispersion bottom and an extended ring in momentum space of large Berry curvature (that we dubbed the `Berry ring of fire', BRF).
In the presence of an isotropic attractive interaction with a parametrically controlled spatial range $\alpha$, we found that a family of broken time-reversal, odd-parity chiral superconductors is generically realized, each characterized by an odd Bogoliubov-de-Gennes (BdG) Chern number $\mathcal{C}$.

With purely local attractive interactions ($\alpha=0$) there is a stark distinction between odd and even number of layers. Odd-$N$ layered systems realize a chiral superconductor with angular momentum $(p_x+ip_y)^N$ and BdG Chern number of $\mathcal{C} = N$. Even-$N$ layered systems, on the other hand, are forbidden from having pairing channels with such purely local attractions, as a consequence of Fermi-Dirac statistics and spatial parity symmetry.
{It is important to note that this forbiddance of chiral pairing channels for even-$N$ layers is special to the contact interaction in the two-band model we consider.}

With spatially extended interactions ($\alpha \neq 0$), pairing channels are stabilized for both odd and even-$N$ layers, with a change of the winding of the superconducting phase occurring at the BRF. For the Fermi surface located 
inside the BRF ($k_F<k_\Omega$), the superconducting phase winds by $\pm 2\pi$, which is the lowest-harmonic odd-parity chiral superconductor ($p_x \pm ip_y$) that benefits from the condensation energy gain. When $k_F$ is
outside the BRF, a whole family of chiral superconducting order parameters can be stabilized at zero temperature, including states with higher odd windings for both even and odd-$N$ layers.
This change is accommodated by the nucleation of an even number of singly quantized momentum-space vortices, which lie (increasingly closer as $\alpha$ increases) on the BRF, to account for the already existing vortex at the origin.
We summarize these findings in Table \ref{tab_summary_conclusions}.

The physical origin of this phase change and the BRF vortices was argued to be the imaginary part of the quantum geometric tensor -- Berry curvature -- which we found to play an analogous role as a real-space magnetic field on a real-space superconducting order parameter.
Interestingly, the phase winding about a closed loop is determined by twice the enclosed Berry flux and a statistical flux factor of $2n\pi$, where $n$ is an odd or even integer to ensure the gap function satisfies Fermi-Dirac statistics and spatial parity symmetry.
The real part of the quantum geometric tensor -- the quantum metric -- provided a spatial extension to nominally short-ranged interactions.
{Though the role of the Berry curvature and the momentum-space flux quantization condition are explicitly derived in the asymptotic limit $\alpha \rightarrow \infty$, these vortices appear even at a finite $\alpha$. Indeed, the mean-field solution of $N$-times phase winding for odd-$N$ layers is smoothly connected to the special $\alpha = 0$ contact limit from analytic continuity.}

\begin{table}[!t]
    \centering
    \renewcommand{\arraystretch}{1.5} 
    \begin{tabular}{|c|c|c|}
    \hline
     & Local interaction & Extended interaction \\
    \hline
    Odd-$N$ & $\mathcal{C} = N$; \hspace{1mm} $\forall k_F$ & 
    \multirow{2}{*}[1ex]{%
    \parbox[c][4.5em][c]{4.6cm}{\centering
    $\mathcal{C} = 
    \begin{cases}
        \pm 1; & k_F < k_{\Omega} \\
        \pm 1, \pm(2n+1); & k_F > k_{\Omega}
    \end{cases}$}} \\
    \cline{1-2}
    Even-$N$ & None; \hspace{3mm} $\forall k_F$ & \\
    \hline
    \end{tabular}
    \caption{Summary of possible chiral superconductors realized in R$N$G based on solutions of the mean-field gap equations, for odd/even-$N$ layers under local ($\alpha = 0$) and extended ($\alpha \neq 0$) range of interactions. ${\cal C}$ denotes the BdG Chern number and corresponds to the total vorticity of the SC order parameter enclosed by the Fermi surface with radius $k_F$.  
    The Berry ring of fire (BRF) has a radius of $k_{\Omega}$, and $n \in \mathbb{Z}>0$.}
    \label{tab_summary_conclusions}
\end{table}

Finally, we discussed how one can tune between these families of chiral superconductors via electron density (which controls $k_F$) or displacement field (which controls $k_\Omega$). 
Chiral superconductors belonging to these families are fully gapped and are separated by critical points which harbor nodal superconductors with gapless excitation spectra{; see also Ref. \cite{yang2024topological} for detailed microscopic studies of topological phase transitions in tetra-layer graphene.} 
We proposed strategies to observe signatures of this behavior experimentally using various established techniques.

A key point that we wish to re-emphasize is that the source of the observed change in the winding of the gap function is the Berry curvature. 
The presence of a concentrated Berry curvature is, however, not a sufficient condition to cause an abrupt change in the gap function winding, as well the number of momentum-space vortices is not always directly related to the magnitude of the Berry flux.
Ultimately, microscopic details -- such as the electronic bandstructure, Fermi surface topology, and pairing mechanism 
-- see Refs. \cite{chou2024intravalley, geier2024chiral, yang2024topological, jahin2024enhanced, qin2024chiral, yoon2025quarter, christos2025finite,parramartinez2025bandrenormalizationquartermetals,roy_2025,roy_2010} for a sample of detailed microscopic studies -- decide whether vortices will be nucleated on the BRF, or whether the preferred gap solution of the family of chiral superconductors examined in this work remains continuous across the inhomogeneous distribution of the Berry curvature.
{Indeed, inclusion of trigonal warping terms reduces the symmetry from $SO(2)\rightarrow C_{3z}$.
This leads to modifying the conduction band dispersion so that $\epsilon(\bfk) \neq \epsilon(-\bfk)$, as well as deforming the BRF into a $C_3$-symmetric shape.
Such reduction in the symmetry can lead to additional pairing channels at finite-momentum (on top of the valley momentum $K$) \cite{sedov2025probingsuperconductivitytunnelingspectroscopy}; nonetheless, for small deformations, it is expected that the BCS pairing we focus on is dominant, since only small portions of a weakly deformed Fermi surface can participate in these additional finite momentum pairings}.
Moreover, the physically-realized pairing channel -- i.e., highest critical temperature -- can also depend on these microscopic parameters (as well as the precise form of the pairing interaction), all of which require more accurate modeling that is beyond the focus of our study; {indeed, there are other competing density wave instabilities that one would also need to consider~\cite{gil2025chargepairdensitywaves}}. Appendix \ref{app_sec_r4g_tc} gives an illustrative demonstration of the competitive temperature evolution of the $N\pm 1$ instabilities in R4G in our model.

For future directions, it would be interesting to investigate this Berry curvature-magnetic field analogy further, with particular interest in devising toy models where the Berry curvature uniformly suppresses the superconducting gap, rather than nucleating vortices -- this would be analogous to the type-I behavior of a superconductor in applied magnetic field.
It will also be interesting to consider finite system geometries to examine the occurrence and stability of the momentum-space vortices to inhomogeneities in the sample, such as strain or applied magnetic field, as well as their effect on real-space properties, such as bulk excitations and edge modes.
{Furthermore, it will be intriguing to understand the effects of trigonal warping more carefully, especially in the context of realizing finite-momentum pairing~\cite{yang2024topological,christos2025finite}, and influence of the Berry curvature on that pairing channel.}
Finally, it will be intriguing to examine the influence of the Bloch wavefunction and quantum geometric properties on the particle-hole exciton condensate, where the long-range Coulomb interaction provides a natural setting to introduce the quantum geometric quantities.

\acknowledgments
We are grateful to Niclas Heinsdorf, Alberto Nocera, Nitin Kaushal, and Sopheak Sorn for insightful discussions.
{We also thank Mathias S. Scheurer and Jason Alicea for helpful comments on the manuscript.}
The work was supported by NSERC, CIFAR, and the Gordon and Betty Moore Foundation’s EPiQS Initiative through Grant No. GBMF11071 at the University of British Columbia (ASP).

\bibliography{references}

\appendix

\section{Parameters of multilayer graphene}
\label{app_miscroscopic_params}
The microscopic tunneling parameters in the kinetic Hamiltonian are: $t_0 = 2.6$eV, $t_1 = 356$meV, $\gamma_0 = \sqrt{3}{t_0}/{2}$, and $v_N = \gamma_0^N/t_1^{N-1}$ \cite{dong2024theory}.
The monolayer graphene lattice constant is taken to be $a = 0.246$nm.
The magnitude of the displacement field employed may appear to be ostensibly large, but is chosen to provide a demarcation of the radius of the BRF and the length scale associated with $\alpha$.
{In particular, the radius of the BRF is $k_{\Omega}=0.23a^{-1}$. We direct the reader to Appendix \ref{app_smaller_d_r5g} for complementary findings at smaller $|D| = 50$ meV.}
The BCS-nature of the superconductor indicates a pairing instability for any small $U$ that is chosen for the various considered cases.
Momentum mesh of $61 \times 61$ and $T= \beta^{-1} = 2 \times 10^{-19}$eV is used unless otherwise stated.

\section{Quantum Geometry of $N$-layer graphene}
\label{app_sec_quantumgeometry}

The quantum geometry provides a means to characterize this eigenvector/band, wherein the quantum geometric tensor is \cite{PARAMESWARAN2013816},
\begin{align}
\eta_{\mu \nu} &= \langle\partial_\mu u | \partial_\nu u \rangle - \langle\partial_\mu u | u \rangle \langle u | \partial_\nu u \rangle = \frac{i}{2} \Omega_{\mu \nu} + g_{\mu \nu},
\label{app_eq_qgt_eq}
\end{align}
where $\Omega_{ab}$ is the anti-symmetrized Berry curvature, and $g_{ab}$ is the quantum metric.
This relation demonstrates how the Berry curvature and the quantum metric are the respective imaginary and real parts of quantum geometric tensor.
Since we are focusing on a two-dimensional system, the relevant Berry curvature component to consider will be $\Omega_{xy} \equiv \Omega$; similarly we will focus on the $g_{xx}, g_{xy}, g_{yy}$ components of the symmetric quantum metric.
We note that in the second equality of Eq.\ \ref{app_eq_qgt_eq} we have defined the anti-symmetrized Berry curvature as $\Omega_{xy} = -i \langle \partial_x u| \partial_y u\rangle + i \langle \partial_y u| \partial_x u\rangle $.

The gauge-dependent Berry connection $\vec{\cala} (\bfk) = -i \bra{u(\bfk)} \nabla_{\bfk} \ket{u(\bfk)}$ is,
\begin{equation}
\begin{aligned}
    \cala^{N}_x(\bfk) &= \frac{N}{2} k_y \frac{\left( - D + \sqrt{v_N^{2}\bfk^{2N} + D^2} \right)}{\bfk^{2} \sqrt{v_N^{2}\bfk^{2N} + D^2}} \\
    \cala^{N}_y(\bfk) &= -\frac{N}{2} k_x \frac{\left( - D + \sqrt{v_N^{2}\bfk^{2N} + D^2} \right)}{\bfk^{2} \sqrt{v_N^{2}\bfk^{2N} + D^2}}.            \label{eq_berry_connection}
\end{aligned}
\end{equation}
Importantly, the gauge-dependent Berry connection is odd in momentum in this gauge choice, $\vec{\cala}(\bfk) = -\vec{\cala}(-\bfk)$.
For a generic gauge choice, $\vec{\cala}(-\bfk) = \vec{\cala}(\bfk) + \nabla_l \chi$, where the second term does not necessarily possess the odd-symmetry.
The corresponding gauge-independent Berry curvature is
\begin{align}
    \Omega^{N} (\bfk) = -\frac{D N^2}{2} \frac{v_N^{2}\bfk^{2(N-1)}}{(v_N^{2}\bfk^{2N} + D^2)^{3/2} } .
\end{align}
Since, the Berry curvature, $\vec{\Omega}(\bfk) = \vec{\nabla}_\bfk \times \vec{\cala}({\bfk})$, is an even function in momentum, this leads to
the conclusion there is always at least one gauge choice where the Berry connection has an odd-symmetry (one such choice is presented in Eq.\ \eqref{eq_berry_connection}).

Similarly, the quantum metric can be easily computed by recalling that it provides a measure of `distance' of the wavefunction in momentum space i.e., $1 - |\langle u (\bfk) | u (\bfk + d \bfk) \rangle|^2 \approx g_{\mu \nu} dk_\mu dk_\nu$,
\begin{align}
    \hspace{-2mm} g^N_{\mu \nu} (\bfk) = \mathcal{G}_{\bfk}
    \begin{pmatrix}
        D^2 + v_N^2 k_y^2 \bfk^{2(N-1)} & - v_N^2k_x k_y \bfk^{2(N-1)} \\
        - v_N^2 k_x k_y \bfk^{2(N-1)} & D^2 + v_N^2k_x^2 \bfk^{2(N-1)}
    \end{pmatrix},
    \label{app_eq_qm}
\end{align}
where $\mathcal{G}_{\bfk} = \frac{v_N^2 N^2 \bfk^{2(N-1)}}{4(v_N^2 \bfk^{2N} + D^2)^2}$.
As seen, both Berry curvature and quantum metric are even functions in momentum $\bfk$.

The quantum metric satisfies the following conditions in relation to the Berry curvature:
\begin{align}
    \sqrt{\det[g^N(\bfk)]} &= \frac{|\Omega^N (\bfk)|}{2} \\
    \text{tr}[g^N(\bfk)] &= \frac{|\Omega^N (\bfk)|}{2} \frac{2 D^2 + v_N^2 \bfk^{2N}}{D \sqrt{D^2 + v_N^2 \bfk^{2N}}}.
\end{align}

\begin{figure}[t]
\includegraphics[width = 0.40\textwidth]{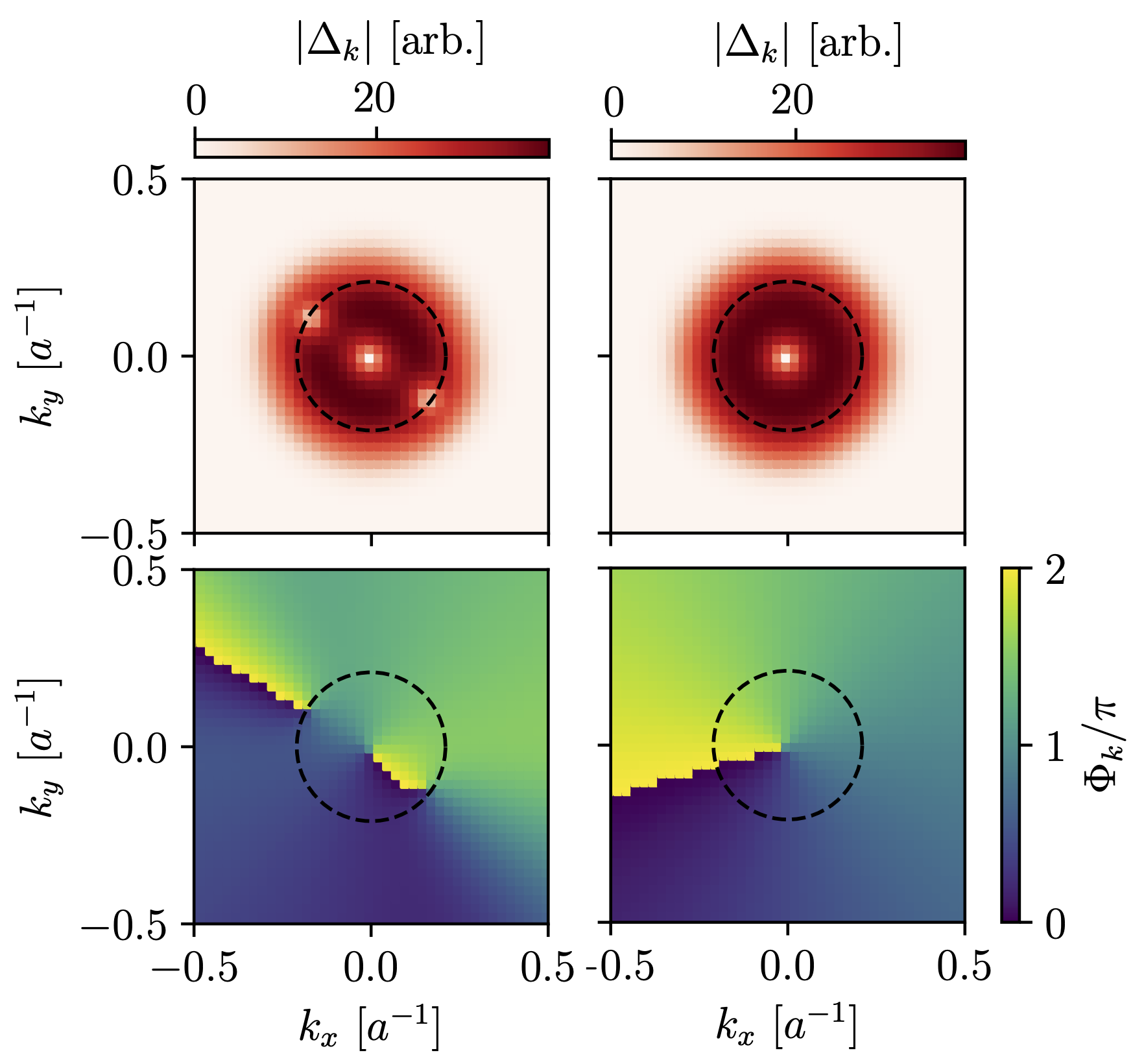}
\caption{Superconducting order parameter of single-phase winding, with its magnitude (top) and phase (bottom) for R7G at $\alpha = 399.4 a^2$. 
The black dashed circle indicates where the Berry curvature is peaked. $U = 10^{-4} a$ eV/m.} 
\label{fig_r7g_single_winding_vortices_or_not}
\end{figure}

\begin{figure*}[t]
\includegraphics[width = 0.9\textwidth]{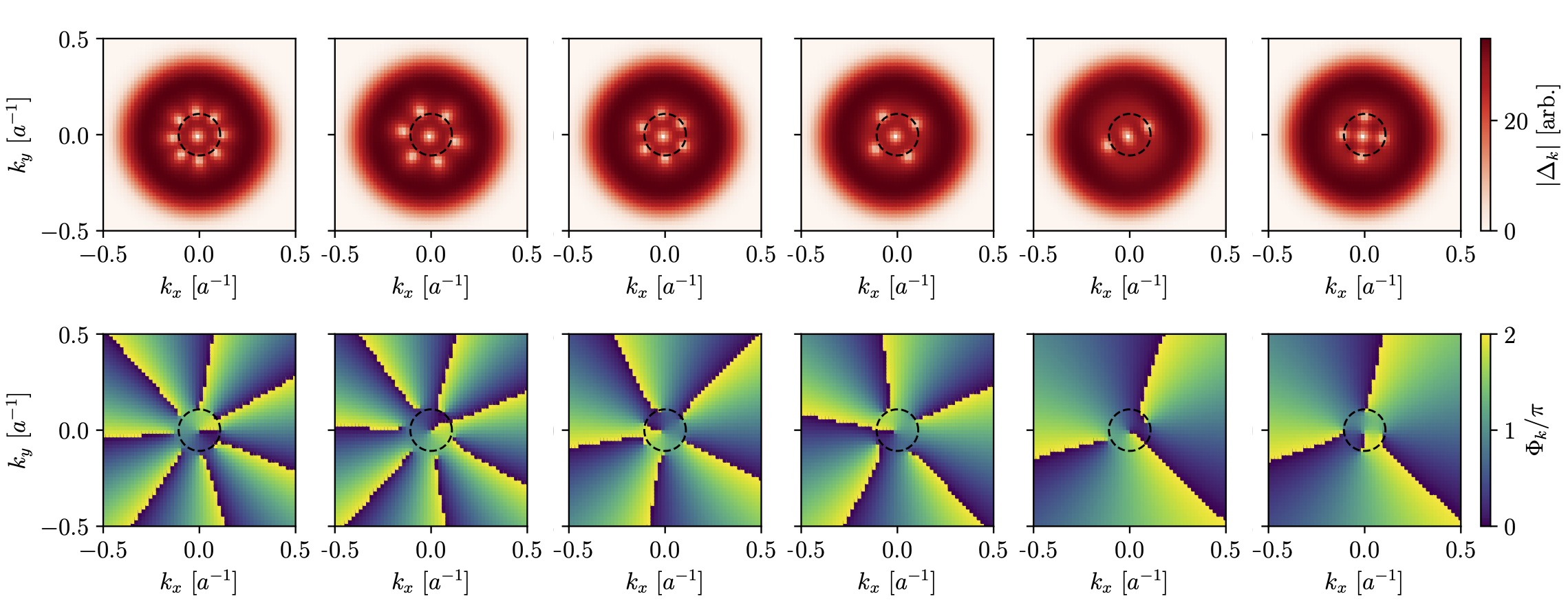}
\caption{Subset of superconducting gap solutions with its magnitude (top) and phase (bottom) for $N=5$ layers at $\alpha = 399.4 a^2$ the attraction length, and $|D| = 50$ meV. 
The black dashed circle indicates where the Berry curvature is peaked. $U = 10^{-4} a$ eV/m, and momentum mesh of $81 \times 81$.} 
\label{fig_r5g_smaller_d}
\end{figure*}

\section{BCS Gap equation Derivation}
\label{app_sec_BCS}

The BCS gap equation is derived from the usual Hubbard–Stratonovich transformation \cite{Coleman_2015, patri2022unconventional}. We begin by rewriting the interaction as,
\begin{align}
\hspace{-3mm} H_{\text{BCS}} =  \sum_{\bfk} \Delta_{\bfk}^{\dag} A_{\bfk} + \Delta_{\bfk} A_{\bfk}^{\dag} + \sum_{\bfk,\bfk'} \Delta^{\dag}_{\bfk'} V^{-1}_{T}(\bfk', \bfk) \Delta_{\bfk},
\end{align}
where $A_\bfk = \psi(-\bfk)\psi(\bfk)$.
Formally integrating out the quadratic fermions in the partition function and taking the logarithm, yields the free energy,
\begin{align}
    F = -\frac{1}{\beta} \sum_{i\omega_n, \bfk} \ln{\left[ \omega_n^2 + \xi_{\bfk}^2 + | \Delta_\bfk|^2 \right]} +\sum_{\bfk,\bfk'} \Delta^{\dag}_{\bfk'} V^{-1}_{T}(\bfk', \bfk) \Delta_{\bfk},
\end{align}
where $\beta = 1/T$ is the inverse temperature. 
The gap equation is obtained by varying the free energy with respect to the order parameter $\Delta_{\bfk}^{\dag}$,
\begin{align}
    \Delta_{\bfk'} = \frac{1}{2} \sum_{\bfk} \frac{\tanh{(\beta E_\bfk/2)}}{E_{\bfk}} V_{T}(\bfk', \bfk) \Delta_{\bfk},
\end{align}
where $E_{\bfk} = \sqrt{\xi_{\bfk}^2 + |\Delta_\bfk|^2}$, and we define $\xi_{\bfk} = \epsilon (\bfk) - \mu$.

\section{Single-phase winding solutions with and without nucleated vortices}
\label{app_r7g_vortex_nucleation}

In this section, we discuss the single-phase winding solution of the superconducting phase.
We present in Fig.\ \ref{fig_r7g_single_winding_vortices_or_not} two stable superconducting gaps that are realizable in R7G for $\alpha = 399.4 a^2$.
As seen, there are two single-phase winding solutions where the chirality of the phase-winding inside and outside the BRF are aligned and opposite.
This entails the nucleation of vortices only when the inner/outer phase windings do not match for the single-phase winding solutions.

\section{Superconducting gap solutions for R5G for smaller displacement field energies}
\label{app_smaller_d_r5g}

We present in Fig.\ \ref{fig_r5g_smaller_d} a subset of the stable superconducting gaps that are realizable in R5G for $\alpha = 399.4 a^2$ and a relatively smaller displacement field energy, $|D| = 50$ meV.
The behaviors are similar to what we obtain and present (in the main text) for larger displacement fields.

\section{Evolution of R5G gap solutions for varied attraction lengths, $\alpha$}
\label{app_r5g_evolution_alpha}

\begin{figure}[t]
\includegraphics[width = 0.45\textwidth]{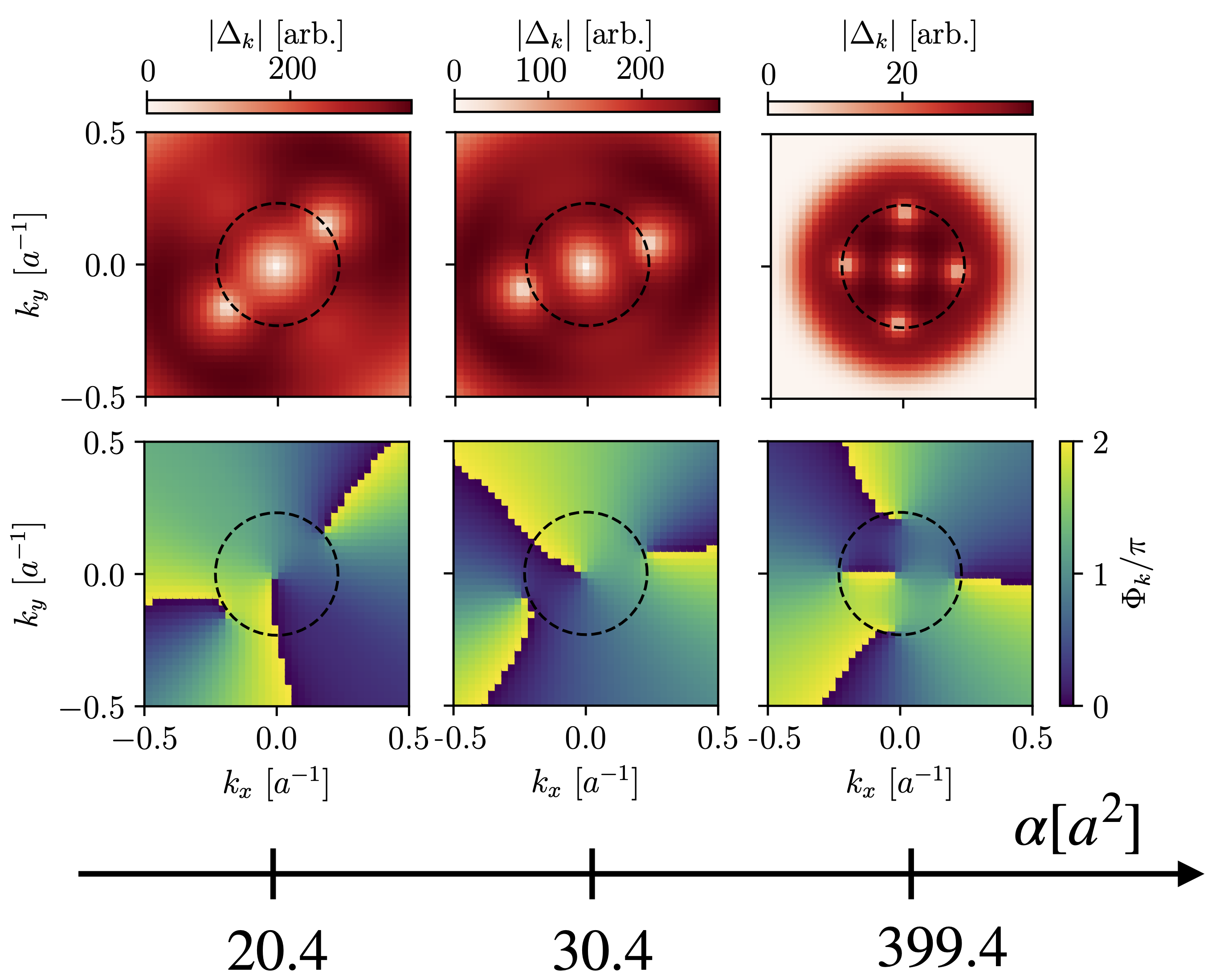}
\caption{Evolution of superconducting order parameter of $N-2$-times phase winding, with its magnitude (top) and phase (bottom) for $N=5$ layers as a function of $\alpha$ the attraction length. 
The black dashed circle indicates where the Berry curvature is peaked. $U = 10^{-4} a$ eV/m.} 
\label{fig_evolution_alpha_N_5_Nm2_winding}
\end{figure}

\begin{figure}[t]
\includegraphics[width = 0.45\textwidth]{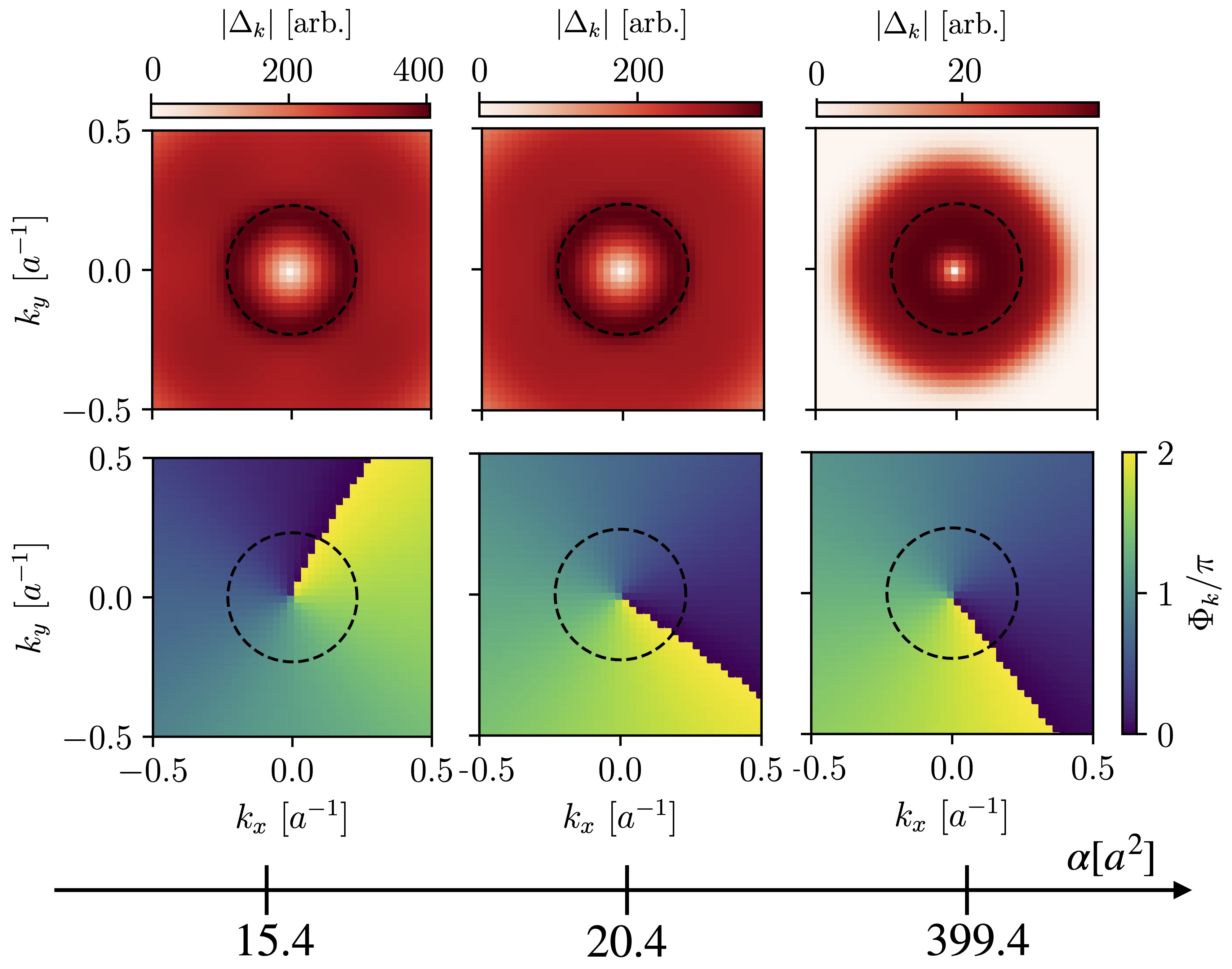}
\caption{Evolution of superconducting order parameter of single-phase winding, with its magnitude (top) and phase (bottom) for $N=5$ layers as a function of $\alpha$ the attraction length. 
The black dashed circle indicates where the Berry curvature is peaked. $U = 10^{-4} a$ eV/m.} 
\label{fig_evolution_alpha_N_5_single_winding}
\end{figure}

We present in Figs. \ref{fig_evolution_alpha_N_5_Nm2_winding} (\ref{fig_evolution_alpha_N_5_single_winding}) the $N-2$ (single-winding) order parameter solutions for R5G for varied $\alpha$.

\section{Generalized winding solutions from higher-order expansion in $\alpha$}
\label{app_higher_order_expansion}

We discuss the expected possible pairing channels that can be stabilized depending on the strength of the non-locality of the attraction. The full expression for the interaction to first order in $\alpha$ is, 
\begin{align}
    \hspace{0mm} V_T(\bfk', \bfk) & \approx 2 U  \mathcal{C}_{\bfk}^2 \mathcal{C}_{\bfk'}^2 v_N^2 \beta_\bfk \beta_{\bfk'} |\bfk|^N |\bfk'|^N e^{iN(\theta_{\bfk'} - \theta_{\bfk})}  \gamma_{\bfk, \bfk'} \delta_{N, \mathbb{Z}_{\text{odd}}}     \label{eq_small_alpha} \nonumber
 \\ & \hspace{0mm} +\alpha U  \mathcal{C}_{\bfk}^2 \mathcal{C}_{\bfk'}^2   |\bfk| |\bfk'| \Big(e^{i (\theta_{\bfk} - \theta_{\bfk'})} + \text{c.c.} \Big) \Big[  \beta_\bfk^2 \beta_{\bfk'}^2 \Big. \nonumber  \\
 & \Big. + v_N^4|\bfk|^{2N} |\bfk'|^{2N} e^{2iN(\theta_{\bfk'} - \theta_{\bfk})} \Big.  \\
   & \Big. + 2\beta_\bfk \beta_{\bfk'} v_N^2|\bfk|^N |\bfk'|^N e^{iN(\theta_{\bfk'} - \theta_{\bfk})} \delta_{N, \mathbb{Z}_{\text{even}}} \Big] \nonumber ,
\end{align} 
where $\gamma_{\bfk, \bfk'} \equiv (1 + \alpha(|\bfk|^2 + |\bfk'|^2))$.
The antisymmetric and symmetric potentials in the limit of $\alpha \rightarrow  0$ are,
\begin{align}
    \hspace{-2mm} V_{A}(\bfk, \bfk') 
    & \approx \sum_{n = 0}^{\infty} (e^{i \Theta_{\bfk,\bfk'}} + e^{-i \Theta_{\bfk,\bfk'}})^{2n + 1} \mathcal{F}^n_{|\bfk|, |\bfk'|}(\alpha), \\
    \hspace{-2mm} V_{S}(\bfk, \bfk') 
    & \approx \sum_{n = 0}^{\infty} (e^{i \Theta_{\bfk,\bfk'}} + e^{-i \Theta_{\bfk,\bfk'}})^{2n} \mathcal{G}^n_{|\bfk|, |\bfk'|}(\alpha),
\end{align}
where $\Theta_{\bfk,\bfk'} = \theta_\bfk - \theta_{\bfk'}$, and ${\mathcal{F}^n_{|\bfk|, |\bfk'|}}(\alpha)$ and ${\mathcal{G}^n_{|\bfk|, |\bfk'|}}(\alpha)$ are functions of the magnitude of the momenta $\bfk  ^{(')}$ and the expansion parameter $\alpha$.
We note that at a given order $\alpha$ there are terms that enhance the phase winding and also benign winding-less terms that are generated. 
We present in Table \ref{tab:alpha_expansion_table} the expected pairing channels in powers of the expansion. 
We note that the higher harmonic $n$ multiple solutions are increasingly suppressed in powers of small $\alpha$ by the weakened pairing potential.
Importantly, there are two familes of pairing channels (last two rows of Table \ref{tab:alpha_expansion_table}) that can be present independent of the parity of $N$.
In particular, the final pairing ansatz which has winding of $\pm(2n+1)=\{\pm1,...\}$ emerges solely due to the $V_A$ term. 
This term (and in particular, the lowest harmonic $n=0$) leads to a gapped odd-parity ($p_x\pm ip_y$) gap, which is a consequence of the Fermi-Dirac statistics.
All these solutions provide an overall odd-times phase winding, as expected for odd-parity superconductors.

\begin{table}[t]
    \centering
    \begin{tabular}{|c|c|}
    \hline
        $N$ & SC Phase Winding  \\
        \hline
        $N_{\rm odd}$ & $N \pm 2n \delta_{\alpha \neq 0}$ \\
        \hline
        $N_{\rm even}$ & $\big[N \pm (2n+1)\big]\delta_{\alpha \neq 0}$ \\
        \hline
        $\forall N$ & $\big[2N \pm (2n+1)\big]\delta_{\alpha \neq 0} $ \\
        \hline
        $\forall N$ & $ \pm (2n+1)\delta_{\alpha \neq 0} $ \\
        \hline
    \end{tabular}
    \caption{Possible pairing channels generated by local ($\alpha$) and non-local ($\alpha \neq0$) pairing interactions. Here $n \in \mathbb{Z}>0$. }
    \label{tab:alpha_expansion_table}
\end{table}

\section{Temperature evolution of subset of R4G's superconducting order parameters}
\label{app_sec_r4g_tc}

\begin{figure}[t]
\includegraphics[width = 0.5\textwidth]{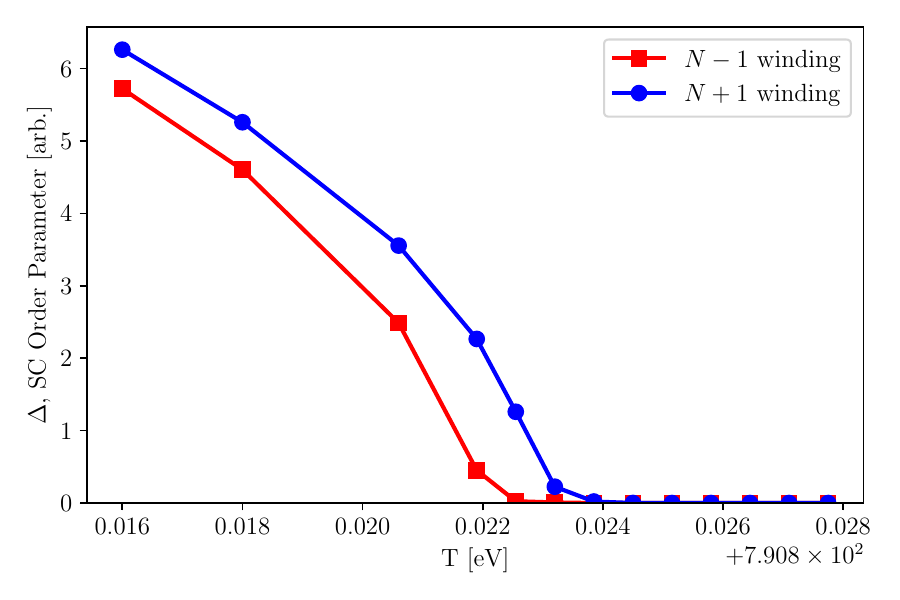}
\caption{Temperature evolution of R4G for $N\pm1$ winding mean-field solutions. Mesh size of $41\times41$, $\alpha = 0.4 a^2$.} 
\label{fig_r4g_0_4_tc_evolution}
\end{figure}

We present in Fig.\ \ref{fig_r4g_0_4_tc_evolution} the temperature evolution for the $N\pm 1$ order parameters for R4G ($N=4)$ for a moderate mesh of 41$\times$41.
It is important to note that determining the precise $T_c$ is not the focus of our study; nonetheless, as an example, we observe that the $T_c$'s are competitive in our model for the particular choice of parameters. 
Perturbations to the model can lead to an enhanced difference in the critical temperatures; we direct the reader to Ref.~\cite{chou2024intravalley}, where the presence of van Hove singularities necessitates a fine momentum mesh.
\\

{
\section{Influence of quantum geometry on magnitude of gap function}
\label{app_sec_qg_gap_influence}

We present the intermediate steps leading from Eq. \eqref{eq_gap_qg} to Eq. \eqref{eq_gap_suppressed_full} to demonstrate the suppression of the magnitude of the gap function at regions of enhanced Berry curvature.
From the flux quantization condition (Eqs. \eqref{eq_closed_loop_flux_quantization}), the complex phases in Eq. \eqref{eq_gap_qg} become $+1$ for $\eta_1 = \eta_2$ and $-1$ for $\eta_1 \neq \eta_2$;  this additional minus sign is compensated by the $\eta_1 \eta_2$ prefactor.
Relabeling the internal momentum $\bfk_1 \rightarrow -\bfk_1$ for the $\eta_1 \neq \eta_2$ cases, leads to the identical momentum integrand for all four combinations of $\{\eta_1, \eta_2\}$ leading to the following gap equation:
\begin{align}
    |\Delta_{\bfk_2}| = U  \sum_{\bfk_1} & \exp{ \Big[ -\left(\alpha \delta_{\mu \nu} + g_{\mu \nu}(\bfk_1) \right) ( \bfk_2 -  \bfk_1)_\mu( \bfk_2 - \bfk_1)_\nu \Big]} \nonumber \\
    &~~~~~~~~ \times \tanh\left({\beta E_{\bfk_1}\over 2}\right)\frac{|\Delta_{\bfk_1}|}{2 E_{\bfk_1}} \label{app_eq_influence}.
\end{align}
Since we are operating in the $\alpha \rightarrow \infty$ limit, $|\bfk_2-\bfk_1| \ll 1$, allowing us to replace $\bfk_1 = \bfk_2 + \bfq$, where $|\bfq|\ll 1/\sqrt{\alpha} \rightarrow 0$.
With this substitution into Eq. \eqref{app_eq_influence}, we expand all quantities to to zeroth order in $\bfq$ i.e., $\Delta_{\bfk_1} \approx \Delta_{\bfk_2} + \mathcal{O}(\bfq^2)$, $E_{\bfk_1} = E_{\bfk_2} + \mathcal{O}(\bfq^2)$, and $g_{\mu \nu}(\bfk_1) \approx g_{\mu \nu}(\bfk_2) + \mathcal{O}(\bfq^2)$.
This leads to a cancellation of the floating factor of $|\Delta_{\bfk_2}|$ on the left and right-side of Eq. \eqref{app_eq_influence} leading to:
\begin{align}
    1 \approx U  \sum_{\bfq} & \exp{ \Big[ -\left(\alpha \delta_{\mu \nu} + g_{\mu \nu}(\bfk_1) \right) \bfq_\mu \bfq_\nu \Big]} \nonumber \\
    &~~~~~~~~ \times \tanh\left({\beta E_{\bfk_1}\over 2}\right)\frac{1}{2 E_{\bfk_1}} +\mathcal{O}(\bfq^2) \label{app_eq_influence_2}.
\end{align}
Finally, since the quantum metric $g_{\mu \nu}(\bfk_1)$ is a semi-definite positive (and $\alpha \rightarrow \infty$), the integrand over $\bfq$ is heavily suppressed for large momenta $\bfq$.
This allows allows us to extend the momentum integration UV bound to $\pm \infty$ for ease of analytical tractability. 
We then perform the standard multi-dimensional Gaussian integral to arrive at Eq. \eqref{eq_gap_suppressed_full} in the main text.
}

\end{document}